\documentclass[12pt]{iopart}
\usepackage{iopams,epsfig}  
\usepackage{epsfig}  
\usepackage{graphicx}
\usepackage{amsmath,amssymb}
\usepackage{bm}
\usepackage{color}

\def\Sqw{S}
\newcommand{\md}[1]{\vert{#1}\vert}
\newcommand{\elem}[3]{\langle{#1}\vert{#2}\vert{#3}\rangle}

\begin{document}

\title[Bragg spectroscopy of the Bose glass phase]{Bragg spectroscopy
  of clean and disordered lattice bosons in one dimension: a spectral
  fingerprint of the Bose glass}

\author{Guillaume Roux$^1$, Anna Minguzzi$^2$, and  Tommaso Roscilde$^3$}
\address{$^1$ LPTMS, Univ. Paris-Sud, CNRS, UMR8626, F-91405 Orsay, France.}
\address{$^2$ Universit\'e Grenoble-Alpes and CNRS, Laboratoire de Physique et Mod\'elisation, des Milieux Condens\'es UMR 5493, Maison des Magist\`eres, B.P. 166, 38042 Grenoble, France}
\address{$^3$ Laboratoire de Physique, CNRS UMR 5672, Ecole Normale Sup\'erieure de Lyon, Universit\'e de Lyon, 46 All\'ee d'Italie, Lyon, F-69364, France}
\ead{\mailto{guillaume.roux@u-psud.fr}, \mailto{minguzzi@grenoble.cnrs.fr}, \mailto{tommaso.roscilde@ens-lyon.fr}}

\begin{abstract}
  We study the dynamic structure factor of a one-dimensional Bose gas
  confined in an optical lattice and modeled by the Bose-Hubbard
  Hamiltonian, using a variety of numerical and analytical approaches.
  The dynamic structure factor, experimentally measurable by Bragg
  spectroscopy, is studied in three relevant cases: in the clean
  regime, featuring either a superfluid or a Mott phase; and in the
  presence of two types of (quasi-)disordered external potentials: a
  quasi-periodic potential obtained from a bichromatic superlattice
  and a box random disorder - both featuring a Bose glass phase. In
  the clean case, we show the emergence of a gapped doublon mode
  (corresponding to a repulsively bound state) for incommensurate
  filling, well separated from the low-energy acoustic mode. In the
  disordered case, we show that the dynamic structure factor provides
  a direct insight into the spatial structure of the excitations,
  unveiling their localized nature, which represents a fundamental
  signature of the Bose glass phase. Furthermore, it provides a clear
  fingerprint of the very nature of the localization mechanism which
  differs for the two kinds of disorder potentials we consider. In special cases, the dynamic
  structure factor may provide an estimate of the position of the
  localization transition from superfluid to Bose glass, in a
  complementary manner to the information deduced from the momentum
  distribution.
\end{abstract}


\submitto{\NJP}

\maketitle

\section{Introduction}

In interacting Bose fluids, the interplay between the effect of
disorder and that of strong interactions displays a rich showcase of
different behaviours. Ultracold atoms offer remarkably the possibility
of exploring such an interplay, since in recent experiments both
disorder and interactions are natural tuning
knobs~\cite{Sanchez-Palencia2010}. In these systems, disorder is
realized either by the application of a speckle potential
\cite{Aspect,Pasienskieta10,Robert-de-Saint-Vincent2010,DeMarco,Jendrzejewski2012,Jendrzejewski2012a}, or by a bichromatic optical lattice composed of
two incommensurate standing waves~\cite{Fallani, Fallani2006, Guarrera2007, Guarreraetal08, LENS, Deissler2009, Deissleretal11, Lucionietal11}, giving rise to a
quasi-periodic (QP) potential~\cite{AubryA79}.
 
In a Bose fluid at zero temperature, Bose-Einstein condensation (or
quasi-condensation in one dimension) occurs generically in the absence
of disorder and in the weakly interacting limit.  We consider two ways
of destabilizing this phase on a lattice: either via Mott localization
due to strong repulsion at commensurate filling; or via Anderson
localization due to strong disorder at any
filling~\cite{Giamarchi1987, Giamarchi1988, Fisher1989}. The resulting
phases - a Mott insulator for strong repulsion, and a Bose glass for
strong disorder (and finite repulsion) - are very different forms of
non-condensed Bose fluids.  Yet, from the point of view of coherence
properties, experimentally probed by time-of-flight measurements in
cold-atom setups, the two phases look similar. In both cases phase
correlations decay exponentially, giving rise to a broad peak in the
momentum distribution. Hence additional information is required for a
direct observation of the Bose-glass phase, which has been long sought
in the context of Bose fluids.

On the other hand, the Bose glass and the Mott insulator are
fundamentally distinguished by the nature of their excitation
spectrum. The lowest energy excitation in a Mott insulator is a
particle-hole excitation with a gap imposed by the energy cost of the
multiple occupation of a lattice site. In the limit of very strong
repulsion, suppressing density fluctuations, such an excitation can be
seen as a free particle/free hole moving on a static background of
particles at integer filling. On the other hand, the lowest
excitations in a Bose glass are \emph{gapless}, and associated with
phonon-like modes localized in rare, locally homogeneous regions of
the sample. Therefore not only the density of states, but the spatial
structure of the excitations provides a distinct fingerprint of the
Bose glass phase with respect to the Mott insulating one.
     
In this respect, the dynamic structure factor, probed by Bragg
spectroscopy in cold-atom experiments \cite{OzeriRMP}, offers the
possibility of characterizing both the spectral density of the
excitations as well as their localized/delocalized nature. 

In the weakly-interacting regime, the dynamic structure factor can be
estimated using the Bogoliubov approach~\cite{Menotti2003}, while at
arbitrary interactions in the 1D uniform system can be obtained from
the Bethe-Ansatz solution of the integrable Lieb-Liniger
model~\cite{CauxCalabrese, Caux2007, Calabrese2007}, the
long-wavelength behaviour in this regime is also captured by Luttinger
liquid theory~\cite{Orignac2012}. In the presence of a lattice, the
dynamic structure factor has been the subject of several analytical
\cite{Huber2007, Golovach2009, Grass2011} and numerical
\cite{Roth2004, Batrouni2005, Rey2005, Pupillo2006, Pippan2009,
  Ejima2011, Ejima2012} studies, as well as of recent experiments
\cite{Clement2009, Du2010, Ernst2010, Fabbri2011,
  Fabbri2012}. In the presence of a disordered potential, only a few
studies have addressed the dynamic response functions of lattice
bosons such as the response for lattice modulation spectroscopy
\cite{Orso2009} and the single-particle spectral function
\cite{Knap2010}, but we are not aware of previous studies of the
dynamic structure factor. In particular the response to lattice
modulation spectroscopy is sensitive to energy only, and it lacks the
momentum resolution which, as anticipated in~\cite{Roux2008} from the
study of the spatial Fourier spectrum of single-particle
excitations, is essential to unveil the localized nature of
excitations.

In this paper, we aim at an extensive investigation of the  dynamic
structure factor in the case of one-dimensional Bose gases in an
optical lattices, and for widely different regimes, encompassing the
weakly interacting limit, the infinitely repulsive case (Tonks limit),
and the regime of intermediate interactions. The two limits of weakly
interacting and infinitely repulsive particles lend themselves to very
convenient theoretical approaches (via Bogolyubov theory and
fermionization, respectively), while the intermediate regime, which is
the most challenging, can be investigated via exact
diagonalization. We particularly focus our attention on the case of a
quasi-periodic potential, featuring a localization transition for a
finite potential depth even in the non-interacting limit - and we
underline the analogies and fundamental differences with respect to a
truly random potential. 

Our main result is that the dynamic structure factor serves as a very
effective diagnostic tool of the localized phases, and, in selected
cases, it might provide a quantitative method to estimate the quantum
phase transition from superfluid to Bose glass, based upon the
localization of the elementary excitations, and therefore
complementary to the analysis of the coherence properties. In
particular, the dynamic structure factor provides clear signatures of the
underlying localization mechanism at play, in all interacting regimes,
and it allows to characterize the Bose-glass phase far beyond its
thermodynamic definition as a compressible and insulating phase.

The paper is organized as follows. Section \ref{s.model} introduces
the model under investigation, \emph{i.e.} the one-dimensional
Bose-Hubbard model in a (quasi-)disordered potential, and the dynamic
structure factor. Section \ref{s.clean} recalls some results on the
limiting cases (weak interaction and infinite interaction) of the clean system, and 
it discusses exact diagonalization results interpolating between these limiting cases.  
Section \ref{s.disordered} studies in details the disordered models: Section \ref{s.Bogo} describes the
dynamic structure factor in the case of weakly interacting bosons,
treated via a Bogolyubov approach, while Section \ref{s.hcbosons}
focuses on the exact solution in the case of hardcore bosons; Section
\ref{s.ED} bridges the two above regimes, contrasting the dynamic
structure factor across the localization transition with the same
quantity across the Mott transition in the absence of
disorder. Section \ref{s.experiments} contains a discussion on the
relevance to experiments while Section \ref{s.conclusions} is
dedicated to conclusions.
 
\section{Bose-Hubbard Hamiltonians in the presence of disorder and methods}
\label{s.model}

\subsection{Models}

We describe one-dimensional bosons in a deep lattice potential and
subject to an external potential using the Bose-Hubbard Hamiltonian,
\begin{eqnarray}
{\cal H} = -J\sum_j [b^{\dag}_{j+1} b_j + \textrm{h.c.} ] + \frac{U}{2} \sum_j n_j(n_j-1) + \sum_j w_j n_j~.
\label{eq:hamiltonian}
\end{eqnarray}
Here $b^{\dag}_j$ is the operator creating a boson at site $j$, $n_j =
b^{\dag}_j b_j$ is the local density, $J$ is the hopping amplitude and
$U$ is the onsite repulsion. We consider lattices with $L$ sites and
$N$ particles, \emph{i.e.} a filling factor $n=N/L$. We denote
the lattice spacing by $a$ in the figures -- when not appearing explicitly in the equations, it is understood that $a=1$. 
The $w_j$ are site-dependent energies which account for both the disorder potential and the
possible presence of a harmonic trapping potential.  For what
concerns the disorder distributions, we will focus on two different
forms of (quasi-)disorder:
\begin{enumerate}

\item A quasi-periodic (QP) potential obtained via a bichromatic
   optical lattice \cite{Fallani}, with the form
 \begin{equation}
w_j = \frac V 2 \sum_j [1 + \cos \left( 2r\pi j + 2\phi\right)]
\end{equation}
with $r$ an irrational number and $\phi$ a random phase-shift on which
averaging can be performed. We choose the experimentally relevant
value $r=830/1076$ \cite{Fallani2006}; when considering periodic
boundary conditions on a lattice of size $L$, we take $r$ as the best
rational approximant in the form $M/L$ (where $M$ is a positive
integer), so that the potential describes a single period over the
entire lattice.

\item A random-box (RB) disorder, for which $w_j$ is a random variable
  uniformly distributed over the interval $[0,V]$.

\end{enumerate}

In both cases, $V/J$ gives the relative strength of the
(quasi-)disorder potential, and it is chosen in such a way that, in
the atomic limit, the (quasi-)disorder closes the Mott gap for $V=U$.
The main difference among the two types of disorder is that, in the
absence of interactions, the QP potential leads to localization of the
single-particle wavefunctions above a critical disorder threshold $V
= 4J$ \cite{AubryA79} while for the RB disorder localization occurs at an
infinitesimal disorder strength. The RB phase diagram was studied
numerically in Refs.~\cite{Batrouni1990, Scalettar1991, Freericks1996,
  Prokofev1998, Rapsch1999}. In the case of the QP potential, the
localization mechanism and the features of the single-particle
wavefunctions and spectrum was extensively studied~\cite{AubryA79,
  Simon1982, Sokoloff85, Thouless1983, Kohmoto1983, Tang1986,
  Hiramoto1992}. In particular, a simple interpretation can be given
in terms of successive band-folding processes (see
e.g. Ref.~\cite{Thouless1983, Barache1994, Roux2008}) or at a
semi-classical level~\cite{Thouless1983, Albert2010}. Bosonization
studies~\cite{Vidal1999, Vidal2001} also show that the quasi-periodic
potential is different from the pure disorder one, in particular, it
does not share the same expected universal Luttinger parameter value
at the localization transition, which was checked in
Ref.~\cite{Roux2008}. The phase diagram of the bichromatic system was
investigated numerically in Refs.~\cite{Roth2003, Roscilde2008,
  Roux2008, Deng2008}.

\subsection{Dynamic structure factor}

The dynamic structure factor $\Sqw(k,\omega)$ is given by the
space-time Fourier transform of the dynamic density-density
correlation function, which for lattice bosons is given by $\langle{
  \delta n_j(t) \delta n_\ell (0)}\rangle$ with $\delta
n_j=n_j-\langle n_j\rangle$. It yields the linear response of the
fluid to a density perturbation transferring a momentum $\hbar k$ and
energy $\hbar \omega$ to the system. Its Lehmann representation reads
\begin{equation}
\Sqw(k,\omega) = \sum_{\alpha\neq 0} \md{\elem{\alpha}{n_k}{0}}^2 \delta(\omega - \omega_{\alpha})\;,
\end{equation}
where $\alpha$ labels the eigenstates of the Hamiltonian ($0$ being
the ground-state), $\omega_{\alpha} = E_{\alpha}-E_0$ are the
excitation energies and
\begin{equation}
n_k = \sum_{j=1}^L e^{i kr_j} n_j\;
\end{equation}
is the density operator at momentum $k$, with $r_j=a(j-L/2)$. On
finite lattice systems with periodic boundary conditions we use $k =
2\pi m/L$ with integer $m$ values. We will also consider the momentum
integrated spectral function $\Sqw(\omega) = \sum_k \Sqw(k,\omega)$.

The dynamic structure factor can be experimentally probed by Bragg
spectroscopy, involving a two-photon transition. Using the
fluctuation-dissipation theorem, it can be extracted from the measured
energy gain of the system per unit time $dE/dt$, according to
\cite{PinesNozieres, Roth2004}
\begin{equation}
\frac{dE}{dt} \propto \omega ~\Sqw(k,\omega)
\end{equation}
where $\omega$ is the frequency difference between the two photons
involved in the transition and $k$ is the wave-vector difference.
Alternatively, the dynamic structure factor can be extracted from the
rate of momentum transfer $dP/dt$ \cite{Brunelloetal2001}
\begin{equation}
\frac{dP}{dt} \propto k ~\Sqw(k,\omega)~.
\end{equation}
Both definitions have been exploited to extract $\Sqw(k,\omega)$ in
recent cold-atom experiments \cite{Veeravalli2008, Ernst2010, Lu2011,
  Clement2009, Fabbri2011, Fabbri2012}.  It is worth mentioning that
the $f$-sum rule allows to compare theory with experiment without
adjustable parameters \cite{Golovach2009}.

Three different theoretical approaches to compute $\Sqw(k,\omega)$ are
used throughout this work and we describe them below.

\subsection{Methods}

\subsubsection{Exact diagonalization}

We computed $\Sqw(k,\omega)$ using the Lanczos algorithm to represent
the low-lying excited states. The method is exact but limited to small
sizes. In the calculation, the maximum number of onsite bosons is
fixed to 6, and 200 iterations are performed to compute the spectral
weights. The delta functions in energy of the discretized excited states are
convolved with lorentzians of width $0.2J$ for $\Sqw(k,\omega)$
and $0.3J$ for $\Sqw(\omega)$. Averaging is performed over 34 samples
with a uniform $\phi$ distribution for the QP potential while 100
samples are used for the RB potential.

\subsubsection{Bogolyubov theory}

For weakly interacting bosons having a finite condensate fraction in
the ground state, it is well established that the ground-state
properties as well as the low-energy excitation spectrum are well
described within Bogolyubov theory. The traditional formulation of
Bogolyubov theory requires the existence of a true condensate, and it
amounts to neglecting terms which are not quadratic in the operators
involving particles out of the condensate. Yet Ref.~\cite{MoraC03} has
shown that an analogous approach to Bogolyubov theory can be applied
to one-dimensional systems featuring only quasi-condensation. Such an
approach is based on a polar decomposition of the Bose operators in
terms of density and phase, $b_j = e^{i\phi_j} \sqrt{n_j}$, and on the
fundamental assumption of weak quantum fluctuations of the density
$\delta n_j = n_j - \rho_j$ around the mean $\rho_j$, as well as weak
quantum fluctuations of the phase difference between neighboring sites
$\theta_j - \theta_{j+1}$. The Hamiltonian can then be expanded in
powers of the density and phase-difference fluctuations around a
reference state, corresponding to non-fluctuating (and vanishing)
phase differences and a classical density profile $\rho_j$ which
satisfies a lattice Gross-Pitaevskii (GP) equation
\begin{equation} 
 -J \left( \sqrt{\rho_{j+1}} + \sqrt{\rho_{j-1}}  \right)  + \left[ U \rho_j - (\mu-w_j) \right] \sqrt{\rho_j} = 0 ~.
\label{e.GPE}
\end{equation}
Here $\mu$ is the chemical potential controlling the number of bosons
in the system (see below) and $\rho_j$ can be identified with the
density profile of the quasi-condensate. Diagonalizing the quadratic
Hamiltonian in the fluctuations amounts to a Bogolyubov transformation
of the density and phase operators to operators $a_s$, $a_s^{\dagger}$
\begin{eqnarray}
\delta n_j & = & \sum_s \left( \delta n_{s,j} ~a_s + \delta n_{s,j}^* ~a_s^{\dagger} \right) + (\partial_N \rho_{j}) ~{\cal P} \nonumber \\
\theta_j &=& \sum_s \left(\theta_{s,j} ~a_s + \theta_{s,j}^*~ a_s^{\dagger} \right) - {\cal Q}
\end{eqnarray}
where ${\cal P}$ and ${\cal Q}$ are canonically conjugated operators
associated with the zero-energy mode, and
\begin{equation}
\delta n_{s,j}  = \sqrt{\rho_j} \left( u_{s,j} + v_{s,j} \right) ~~~~~~~~~ \theta_{s,j} = \frac{ u_{s,j} - v_{s,j}}{2i\sqrt{\rho_j}}~.
\end{equation}
% The Hamiltonian governing the quadratic quantum fluctuations reads ${\cal H}_2 = \sum_s \omega_s b_s^{\dagger} b_s + \frac{{\cal P}^2}{\mu_0 
The $u_{s,j}$, $v_{s,j}$ amplitudes satisfy the Bogolyubov-de Gennes
(BdG) equations
\begin{equation}
 {\cal L}  \begin{pmatrix} |u_s\rangle \\ |v_s\rangle \end{pmatrix} = \begin{pmatrix} {\cal H}_2 & {\cal H}_U \\ -{\cal H}_U & - {\cal H}_2 \end{pmatrix}  
 \begin{pmatrix} |u_s\rangle \\ |v_s\rangle \end{pmatrix} = \epsilon_s \begin{pmatrix} |u_s\rangle \\ |v_s\rangle  \end{pmatrix}
\label{e.BdG}
\end{equation}
where 
\begin{eqnarray}
 {\cal H}_2 & = &  -J \sum_j \left( |j\rangle \langle j+1 | + {\rm h.c.} \right ) + \sum_j \left( 2U \rho_j + w_j - \mu \right) |j\rangle \langle j | \nonumber \\
 {\cal H}_U & = & \sum_j U \rho_j |j \rangle \langle j |
\end{eqnarray}
and $|u_s\rangle = \sum_j u_{s,j} |j\rangle$, $|v_s\rangle = \sum_j
v_{s,j} |j\rangle$. The properties of the non-Hermitian eigenvalue
problem of Eq.~\eqref{e.BdG} are well known \cite{BlaizotR84}; in
particular the solutions of the BdG equations with non-zero energy
$\epsilon_s$ satisfy the normalization condition
\begin{equation}
\langle u_s | u_s' \rangle - \langle v_s | v_s' \rangle = \delta_{ss'}~.
\label{e.norm}
\end{equation}
In the absence of disorder the $u$ and $v$ modes are plane waves, and
the corresponding energies, labeled by momentum ($s\to k$), have the
well-known form
\begin{equation}
\omega_k = \sqrt{ e_k (e_k + 2nU)}\;,\quad\text{with}\quad e_k = 4J \sin^2(k/2)\;.
\end{equation}
Given that quantum fluctuations in the density are linear in the
$a_s$, $a_s^{\dagger}$ operators, within the quadratic approximation
for quantum fluctuations the quantum corrections to the density
profile vanish (which is consistent with the image of $\rho_j$ as the
mean local density). Therefore the total particle number is given by
$N = \sum_j \rho_j$, and the chemical potential $\mu$ in
Eq.~\eqref{e.GPE} is fixed so as to impose the desired lattice filling
at the level of the GP equation.
  
The Bogolyubov theory for quasi-condensates of Ref.~\cite{MoraC03} has
been applied to 1D Bose gases in a disordered potential in a number of
recent papers
\cite{Fontanesietal09,Fontanesietal10,LuganSP11,GaulM11}. In
particular Refs.~\cite{Fontanesietal09,Fontanesietal10,FontanesiPhD}
provide an explicit expression for the one-body density matrix,
$g^{(1)}(j,l) = \langle b_j^{\dagger} b_l \rangle$ in terms of the
solution of the GP and BdG equations, reading
\begin{equation}
g^{(1)}(j,l) = \langle b_j^{\dagger} b_l \rangle = \sqrt{\rho_j \rho_l} \exp\left[-\frac{1}{2} \sum_s \left(\frac{v^{\perp}_{s,j}}{\sqrt{\rho_j}} - \frac{v^{\perp}_{s,l}}{\sqrt{\rho_l}}\right)^2 \right] 
\end{equation}
where $u_{s,j}^{\perp}$, $v_{s,j}^{\perp}$ are the coefficients of the
vectors $|u_s\rangle$ , $|v_s\rangle$ orthogonalized with respect to
the quasi-condensate
\begin{equation}
|u^{\perp}_s\rangle = |u_s\rangle - \langle \psi_0 |u_s\rangle | \psi_0\rangle
\end{equation}
(and similarly for $|v^{\perp}_s\rangle$). Here $|\psi_0 \rangle =
\sum_j \psi_{0,j} |j\rangle$ is the normalized quasi-condensate mode,
with $\sqrt{\rho_j} = \sqrt{N} \psi_{0,j}$.
 
The calculation of the dynamic structure factor of a weakly
interacting Bose gas has been addressed within Bogolyubov theory in
various references \cite{WuGriffin97,Zambellietal00}. In the case of
quasi-condensates its expression turns out to be analogous to that of
conventional Bogolyubov theory \cite{Zambellietal00}, namely
\begin{equation}
\Sqw(k,\omega) = \sum_s |\delta\tilde\rho_s(k)|^2 \delta(\omega - \epsilon_s)  
\label{e.SkomBog}
\end{equation}
where the form factors $\delta\tilde\rho_s(k)$ read
\begin{equation}
\delta\tilde\rho_s(k) = \sum_j e^{-ikr_j} \sqrt{\rho_j} (u_{s,j} + v_{s,j})~. 
\label{e.rhosk}
\end{equation}
Notice that here we restrict our attention to the case $\omega>0$, so
that the zero-mode contributions disappear from $\Sqw(k,\omega)$.  In
the absence of an external potential, the dynamic structure factor is
a $\delta$-peak resonance at the Bogolyubov mode $\epsilon_k$~\cite{Menotti2003}:
\begin{equation}
\Sqw(k,\omega) = \frac{e_k}{\omega_k} ~N \delta(\omega-\omega_k)~.
\label{e.sqw-bogo}
\end{equation}
   
The Bogolyubov theory for quasi-condensates is quantitatively
consistent as long as the quantum fluctuations of the density and
relative phase remain weak. In particular, the phase-density
formulation requires the condition $\rho_j \gg 1$ to be fulfilled in
order for the phase operator to be well defined as a (quasi-)Hermitian
operator \cite{MoraC03}. Moreover, a large quasi-condensate density on
each site is also necessary for the relative particle fluctuations to
be small, since $\langle (\delta n_j)^2 \rangle / \rho^2_j \geq
1/\rho_j$ \cite{MoraC03}. As we will see, this condition will strongly
limit the range of validity of our results.
  
From the practical point of view, we numerically solve the
Gross-Pitaevskii equation via split-operator imaginary-time
propagation, and the Bogolyubov-de Gennes equations by diagonalization
of the non-Hermitian ${\cal L}$ matrix using the LAPACK libraries, as
described in previous references \cite{Castroetal06}. We present
results for lattices with $L=256$ and $L=512$ sites. Unless otherwise
specified, the results for QP potentials are averaged over $\sim 50$
values of the spatial phase $\phi$ of the potential.

\subsubsection{Hardcore-boson limit}

While Bogolyubov theory applies to weakly interacting bosons at large
filling, we can also consider the opposite limit of infinitely
repulsive bosons, $U\to\infty$, and low filling $n<1$. This limit
corresponds to the 1D Tonks-Girardeau gas of hardcore bosons (HCB), in
which the forbidden double occupancy of the sites can be incorporated
in a redefinition of the bosonic operators, $b_i \to \tilde{b}_i$
satisfying bosonic commutation relations offsite and anticommutation
relations onsite, $[\tilde{b}_j, \tilde{b}_l^{(\dagger)}] =0$ for
$j\neq l$ and $\{\tilde{b}_j, \tilde{b}_j^{\dagger}\} = 1$, $\{
\tilde{b}_j, \tilde{b}_j\} = \{\tilde{b}_j^{\dagger},
\tilde{b}_j^{\dagger}\}=0$. The hardcore boson operators
$\tilde{b}_j^{(\dagger)}$ can be transformed to fermionic ones
$c_j^{(\dagger)}$ via a Jordan-Wigner transformation \cite{LiebSM61},
mapping exactly the hardcore boson Hamiltonian to free fermions with
chemical potential $\mu$
\begin{equation}
{\cal H} = -J \sum_j \left( c_j^{\dagger} c_{j+1} + {\rm h.c.} \right) + \sum_j (w_j - \mu) c^{\dagger}_j c_j
\end{equation}
which lends itself to efficient exact diagonalization. In the
following we will focus on systems with open boundaries or in a trap,
and therefore we omit in the Hamiltonian the boundary terms arising
from the non-local nature of the Jordan-Wigner transformation.

Most notably, the hardcore boson density coincides with the fermionic
one, $\tilde{b}_j^{\dagger} \tilde{b}_j = c^{\dagger}_j c_j$, so that
the dynamic structure factor of the hardcore bosons corresponds to
that of the free fermions, taking the simple expression
\begin{equation}
\Sqw(k,\omega) = \sum_{\alpha\beta} |\rho_{\alpha\beta}(k)|^2 ~f(e_{\beta},T) [1-f(e_{\alpha},T)]~ \delta(\omega -\omega_{\alpha\beta})
\label{e.SkomHCB}
\end{equation}
where $\omega_{\alpha\beta} = e_\alpha - e_\beta$, $e_\alpha$ are the
eigenenergies of the single particle problem in the QP potential,
$f(e,T) = \{\exp[(e-\mu)/(k_B T)]+1\}^{-1}$ is the Fermi-Dirac
occupation factor at a finite temperature $T$, and
\begin{equation}
\rho_{\alpha\beta} (k) = \sum_j e^{ikr_j} \psi^*_{\alpha,j} \psi_{\beta,j}~.
\label{e.rhoab}
\end{equation}
Hence we observe that, for hardcore bosons, the $k$-dependence of the
dynamic structure factor describes the power spectrum in momentum space
of the overlap function $\psi_{\alpha j}^{*} \psi_{\beta j}$ between
occupied and unoccupied single-particle states, connected by the energy
transfer $\hbar \omega$.

\section{Dynamic structure factor of the clean system: superfluid and Mott phases}
\label{s.clean}

The phase diagram of the clean Bose-Hubbard model displays two
phases~\cite{Fisher1989}: the superfluid phase (SF) which occurs
generically at incommensurate densities, and the Mott-insulator (MI)
phase which occurs only at commensurate fillings and beyond a critical
interaction strength $U_c$ (for a one-dimensional system with filling
$n=1$, $U_c \simeq 3.3J$). In the incommensurate case at densities $n
< 1$, there are two main types of low-energy excitations contributing
to the dynamic structure factor in the Bose-Hubbard model. The first
type is represented by gapless acoustic modes related to the
superfluid regime and which have, in the long wavelength limit, the
dispersion relation $\omega(k) \simeq c k$, with $c$ the sound
velocity. The second type is represented by doublon excitations,
namely repulsively bound states of two particles occupying the same
site, occurring when the repulsion energy exceeds the bandwidth; these
bound states appear at the two-body level, and survive in the
many-body case~\cite{Winkler2006}. Its energy creation cost is about
$4J+U$ in the strong coupling limit. In the Mott regime, the
elementary excitations form a particle-hole continuum. This gapped excitations of the Mott phase have a typical
energy cost of the order of $U$ and the shape of the particle-hole continuum is known in the deep Mott limit
~(see e.g. Ref.~\cite{Golovach2009}). This excitation is
essential in understanding the dynamical properties of the system at
large $U$, and it is the main excitation in the Mott phase where sound
modes are absent.

\begin{figure}[t]
\centering
\includegraphics[width=\textwidth,clip]{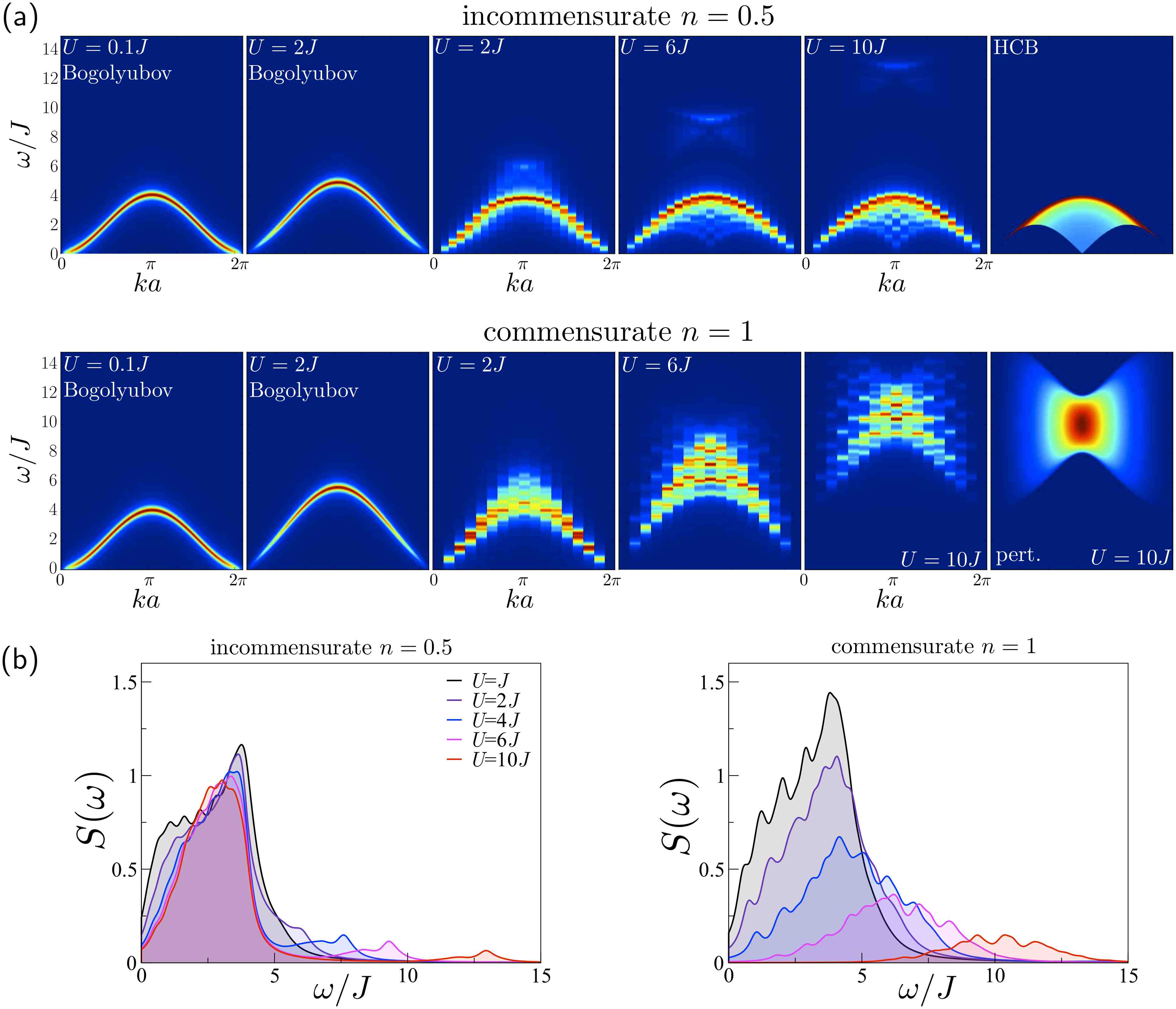}
\caption{\textsf{(a)} Dynamic structure factor $\Sqw(k,\omega)$ for
  the clean system. Upper panel: for an incommensurate filling $n=0.5$ ($L=22$)
  from weak to strong interactions. Bogolyubov and hardcore-boson (HCB)
  results are compared to exact diagonalization. Lower panel: for a
  commensurate filling $n=1$ ($L=16$) where the Mott phase emerges at
  large $U/J$. The perturbative (pert.) result is also shown. Color scales from
  different figures are different. \textsf{(b)} Integrated dynamic
  structure factor $\Sqw(\omega)$, allowing to compare the spectral
  weights for different $U/J$ for the two situations.}
\label{fig:clean-system}
\end{figure}

These elementary excitations are visible in the dynamic structure
factor of the clean Bose-Hubbard model. We give in
Fig.~\ref{fig:clean-system}(a) the evolution of the dynamic structure
factor $\Sqw(k,\omega)$ for increasing interaction $U/J$ and for two
typical densities: incommensurate ($n=0.5$) and commensurate ($n=1$).

We start with the incommensurate case. At small $U/J$, we display the
Bogolyubov result of Eq.~\eqref{e.sqw-bogo}. The overall behavior has
a form of an arc on $[0,2\pi]$ due to the periodicity in momentum
space.  For $U=2J$, we see that the Bogolyubov result and the exact diagonalization
result differ qualitatively. While it is expected that Bogolyubov
theory fails to account for such a relatively strong interaction, we
observe that the essential difference emerges around $k\sim \pi$ where
local physics is dominant. The spectrum is there split into two lines
which we interpret as an hybridization between the acoustic modes and
the gapped doublon state. Then, the peak emerging on top of the
acoustic branch is attributed to the doublon state. This doublon mode
is roughly centered around the energy $4J+U$, as we see for increasing
$U/J$, and as expected for repulsively bound
pairs~\cite{Winkler2006}. When its energy increases, its spectral
weight decreases, as one can see from $\Sqw(\omega)$ plotted in
Fig.~\ref{fig:clean-system}(b). In addition, increasing the
interaction strength transfers spectral weight to the low-energy
states at the $k=2k_F$ wave vector, $k=2\pi n$ (\emph{i.e.} $k = \pi$ in
the figure) corresponding to back-scattering (see eg
\cite{Imambekov2012}). This is the very analogue of what is found in
the Lieb-Liniger model \cite{CauxCalabrese} in the absence of the
lattice. In the hardcore boson limit, the spectrum corresponds to the
XX spin chain model which displays the famous Pearson-De Cloiseaux
continuum (see e.g. \cite{Cloizeaux1962, Golovach2009}). In this
limit, the doublon excitation is no longer part of the spectrum. Thus,
exact diagonalization nicely interpolates between the two regimes, and
it exhibits the evolution of the doublon excitation in the
spectrum when approaching the HCB limit.

Looking at the commensurate $n=1$ case in Fig.~\ref{fig:clean-system},
one can see the transition from the SF to the MI phase when
interactions are increased, although the opening of the gap in the
dispersion relation is appreciable only at $U$ sizably larger than $U_c$
because of finite-size effects (the gap opens exponentially
slowly). One switches from the SF behavior at $U=J$, similar to the
incommensurate case and in qualitative agreement with Bogolyubov
theory, to a fully gapped excitation spectrum in the MI regime which
corresponds to the particle-hole dispersion centered around $\omega
\sim U$. For $U=2J$, there is a single peak in the acoustic mode, with
a short lifetime at $k=\pi$ compatible with previous results for the
same filling~\cite{Pippan2009, Ejima2011, Ejima2012}. Interestingly,
in the intermediate interaction regime $U\sim 6J$, the system is
gapped but the particle-hole dispersion has a significant weight at
the lowest frequencies, reminiscent of the acoustic mode
spectrum~\cite{Grass2011}.  According to perturbation
theory~\cite{Golovach2009}, for large enough $U/J$ the spectrum is
predicted to display a butterfly-like shape with a maximum weight
around $k=\pi$. For $U=10J$, where the gap is sizeable, $S(k,\omega)$
has a similar support as the perturbative prediction of
Ref.~\cite{Golovach2009}, but the weight distribution is not yet
symmetric around $\omega=U$. We observe that ED can quantitatively
covers the evolution of $S(k,\omega)$ from the weakly to the strongly
interacting regime, and that Bragg spectroscopy can capture the
opening of the gap. Yet, in realistic conditions with a trap, one
would have to consider the effect of the inhomogeneity of the
system~\cite{Golovach2009}.

\section{Dynamic structure factor for the quasi-periodic system}
\label{s.disordered}

The impact of disorder on elementary excitations of interacting bosons
can be expected to be qualitatively similar to that on single-particle
states, leading in particular to localization of the spatial support
of the excitation modes. The connection between single-particle and
many-body physics is evident within the Bogolyubov and HCB approaches,
since the spatial structure of the excitation modes comes from the
solution of the single-particle Schr\"odinger's equation (for HCB) or
the solution of BdG equations in the presence of a (quasi-)disordered
potential. Yet the same connection is far less obvious in the full
Bose-Hubbard model. In what follows we will first describe our results
in the Bogolyubov and HCB regimes, and then show how exact
diagonalization allows to interpolate between the above regimes. In
the case of exact diagonalization we also compare the case of QP and
RB potentials, and we show that, due to the different nature of the
localization mechanism at play, these two potentials lead to very
distinct features in the dynamic structure factor.

\subsection{Results from Bogolyubov theory}
\label{s.Bogo}
 
We present here the results for the dynamic structure factor of
one-dimensional weakly interacting bosons in a quasi-periodic
lattice. We begin our discussion with the non-interacting limit, which
serves as a useful reference for the results in the interacting case.

\begin{figure}[t]
\centering
\includegraphics[width=\columnwidth,clip]{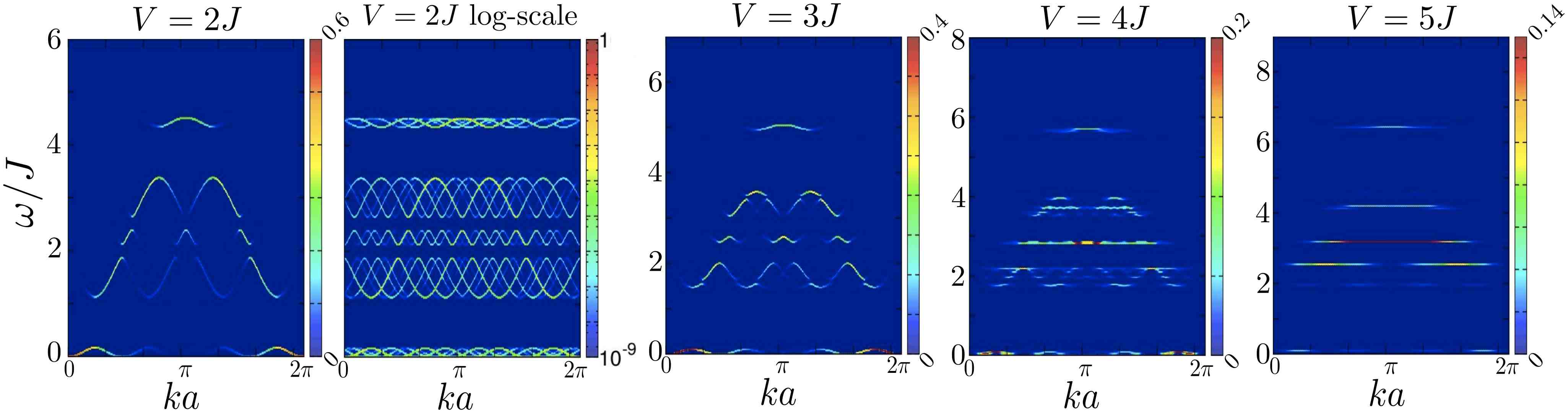}
\caption{Dynamic structure factor $\Sqw(k,\omega)$ for free bosons in
  a quasi-periodic potential for increasing strength $V/J$ of the
  disorder potential. The case $V=2J$ is shown in log-scale to highlight the
  underlying excitations bands.}
\label{fig:SqomU0}
\end{figure}

\subsubsection{U=0}
\label{s.U0}
  
In the case of an ideal gas, the dynamic structure factor takes the
simple form
\begin{equation}
\Sqw(k,\omega) = \sum_{\alpha} \left |\rho_{\alpha 0}(k) \right|^2 \delta(\omega-\omega_\alpha)
\end{equation}
where $\sum_{\alpha}$ runs over the single-particle eigenstates,
$\omega_\alpha = e_\alpha- e_0$ is the excitation energy of the
$\alpha$ state, and $e_\alpha$ is the single-particle eigenenergy
corresponding to the lattice eigenfunction $\psi_{\alpha,j}$;
$\rho_{\alpha 0}(k)$ is the Fourier transform of the overlap function
between the ground-state and the $\alpha$-th excited state, defined in
Eq.~\eqref{e.rhoab}.

If the ground-state is close to a $k=0$ plane wave, the
$\rho_{\alpha0} (k)$ form factor is essentially proportional to the
Fourier transform of the excited state $\psi_{\alpha}$. As a
consequence $\Sqw(k, \omega)$ gives the power spectrum in momentum
space for the excited state at energy $\epsilon_\alpha =
\omega$. Therefore, even in the absence of translational invariance
(broken by the QP potential), the presence of a sharp ridge in
$\Sqw(k,\omega)$ gives an effective energy-momentum dispersion
relation for the excited states of the system.
 
Fig.~\ref{fig:SqomU0} shows the single-particle dynamic structure
factor for an increasing strength of the QP potential.  We observe
that the $e_{\alpha} = e_k$ dispersion relation of free particles in
the lattice, characterized by a single cosine band, breaks up into
sub-bands for a finite $V$. We shall first focus on the delocalized
phase $V< V_c$. The appearance of the sub-bands can be related with
the fact that the QP potential introduces a quasi-periodic structure
in the lattice, whose spatial period corresponds to the period of the
beating between the underlying lattice and the incommensurate
potential, $l_{\rm QP} = (1-r)^{-1}$. Correspondingly features in momentum
space appear at the edges of a (pseudo-)Brillouin zone with
width $k_{\rm QP}= 2\pi(1-r)$, namely at $k_{\rm QP}/2$, $\pi\pm
k_{\rm QP}/2$, etc.; these are indeed the (approximate) momentum
locations at which the gaps between the sub-bands appear in
$\Sqw(k,\omega)$. Within each subband the excitations are delocalized
with sharp momentum content, and they exhibit a cosine-like dispersion
with the periodicity of the pseudo-Brillouin zone. Yet, due to the
incommensurability, the pseudo-Brillouin zone cannot fill the
Brillouin zone of the underlying lattice an integer number of times,
and hence the sub-bands dispersion curves essentially fade away in
$\Sqw(k,\omega)$ after a few periods (in fact a closer inspection
shows that they persist over the whole Brillouin zone, and they even
wind around it giving rise to a very complex pattern, which
nonetheless is only seen in logarithmic scale -- see
Fig.~\ref{fig:SqomU0} for $V=2J$). This fading dispersion relation can
be understood within a perturbative picture for the QP potential: a
particle with momentum $k$ and energy $e_k$ (in the absence of the QP
potential) is scattered by the QP potential and it can acquire a
momentum $p~k_{\rm QP}$ at $p$-th order in perturbation theory, but
due to the incommensurability there is no finite order in perturbation
theory which can connect the initial state to a resonant unperturbed
state, and therefore the particle remains ``localized'' around its
initial momentum $k$ with fading components at $k \pm k_{\rm QP}$, $k
\pm 2k_{\rm QP}$, etc. This picture of localization in momentum space
is valid beyond perturbation theory, and it relies on the exact
duality of the non-interacting model under Fourier transformation
\cite{AubryA79, Sokoloff85}.
    
For $V>V_c$, $\Sqw(k,\omega)$ undergoes a radical change: the
dispersive nature of the excitations within the subbands disappears,
and $\Sqw(k,\omega)$ acquires features which are very broad in
momentum space, while retaining a sharp nature in the frequency
domain. This corresponds to the appearance of strongly localized
modes, possessing a large uncertainty in momentum space. The large
broadening of the structure of $\Sqw(k,\omega)$ in momentum space is
therefore the signature of localization, and it will reappear as a
leitmotiv in the analysis of the results for the interacting system.

\begin{figure}[t]
\centering
\includegraphics[width=0.6\textwidth,clip]{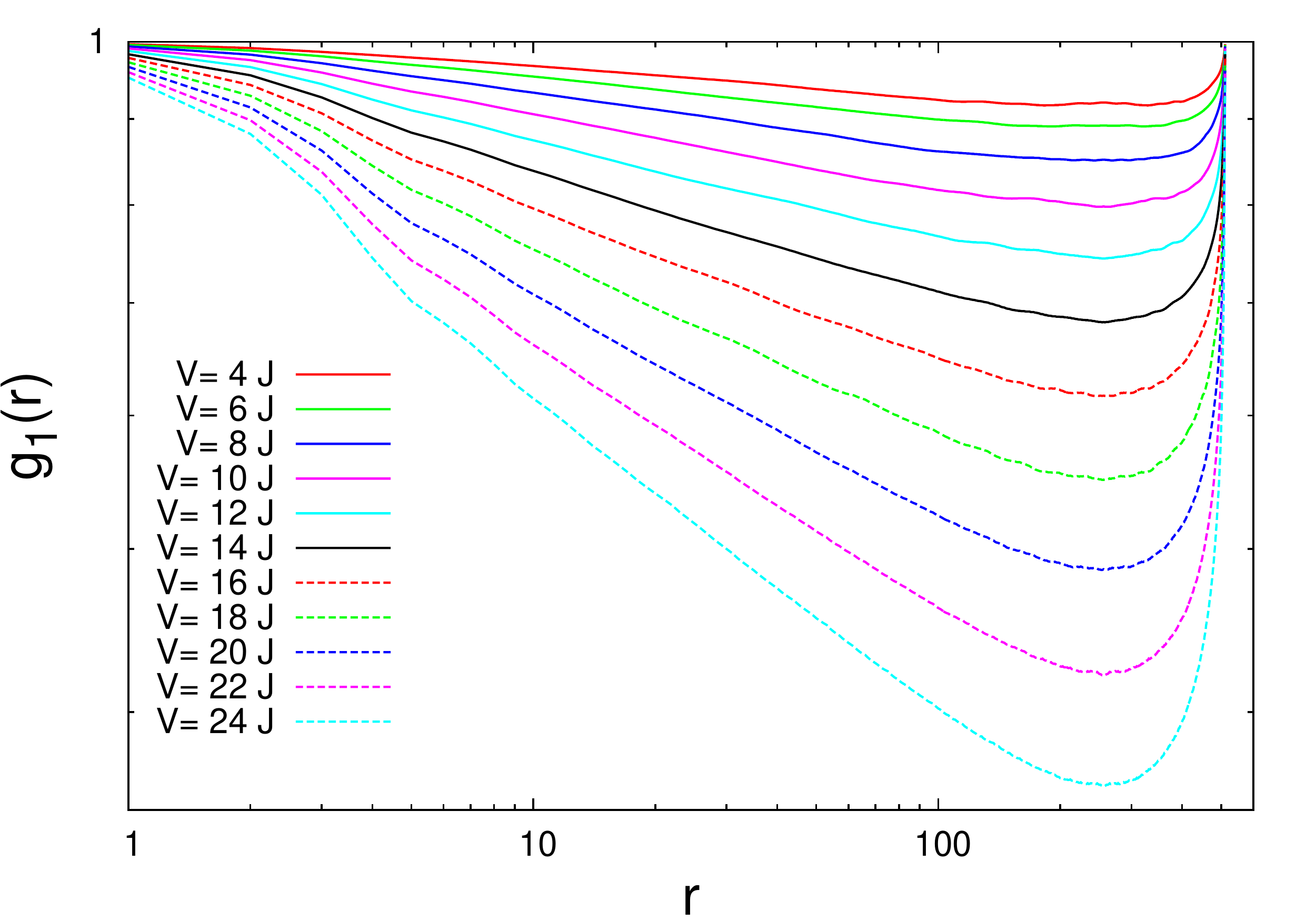}
\caption{One-body density matrix for the 1D weakly interacting lattice
  Bose gas in a QP potential of increasing strength. The parameters
  are $r = 395/512$, $U = 0.1 J$, $n=10$, and $L=512$.}
\label{fig:g1}
\end{figure}

\subsubsection{$U>0$.}
\label{s.Bogo.int}
In the following we present our results for the weakly interacting
case. We will mostly present results for $U = 0.1 J$ and a lattice
filling $n = 10$ to satisfy the conditions of validity of Bogolyubov theory -- although
we have also investigated the interaction strengths $U = 0.01 J$ and
$0.5 J$, displaying similar features to $U=0.1 J$. As already discussed in
Refs.~\cite{CastinM03,Fontanesietal09,Fontanesietal10,CetoliL10} for
the case of bosons in continuum space, Bogolyubov theory is capable of
describing quasi-condensates, and specifically a power-law decaying
one-body density matrix. This is also verified for a lattice system
and in the presence of a QP potential, as shown in
Fig.~\ref{fig:g1}. In particular we observe that weak interactions
make the quasi-condensate state robust to the QP potential, and they
promote it to values of $V$ well beyond the critical value $V_c=4 J$
for the non-interacting system. In the case of a 1D gas in continuum
space and subject to a speckle or quasi-periodic potential,
Refs.~\cite{Fontanesietal09,Fontanesietal10,CetoliL10} have shown that
Bogolyubov theory allows to quantitatively describe the localization
transition in one dimension in the presence of interactions; such a transition is detected by the appearance
of an exponential decay in the one-body density matrix. This has
allowed the authors of
Refs.~\cite{Fontanesietal09,Fontanesietal10,CetoliL10} to track the
interaction-induced shift of the critical disorder strength. In the
lattice system under investigation, on the other hand, we rather find
that Bogolyubov theory fails to reproduce quantitatively this
transition. Indeed, for all the interaction strengths we considered,
we find that a quasi-condensate phase is observed over the whole range
of applicability of the theory, namely for disorder strengths $V$
which do not lead to excessive fragmentation of the density
profile. Indeed, if the disorder strength is too large, there will
appear sites in the lattice with $\rho_i \ll 1$, clearly violating the
condition of weak density fluctuations. We do observe a change of the
$g^{(1)}$ function from an algebraic to an exponential decay with
increasing $V$, but this occurs at unrealistically large values of
$V$, well beyond the value $\sim V_c + Un$ which na\"ively represents
the critical value for a QP potential screened by the interactions,
and well beyond the range of validity of the theory.

\begin{figure}[t]
\centering
\includegraphics[width=\textwidth,clip]{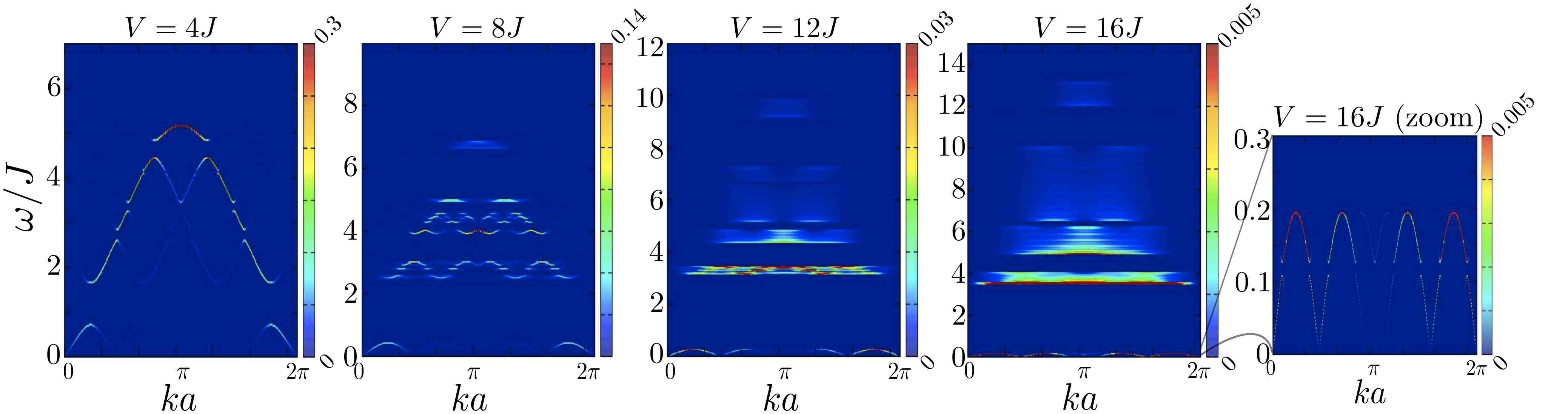}
\caption{Dynamic structure factor $\Sqw(k,\omega)/N$ for the 1D weakly
  interacting lattice Bose gas in a QP potential of increasing
  strength. Same parameters as in Fig.~\ref{fig:g1}.}
\label{fig:SqomU0.1}
\end{figure}

\begin{figure}[t]
\centering
\includegraphics[width=0.6\textwidth,clip]{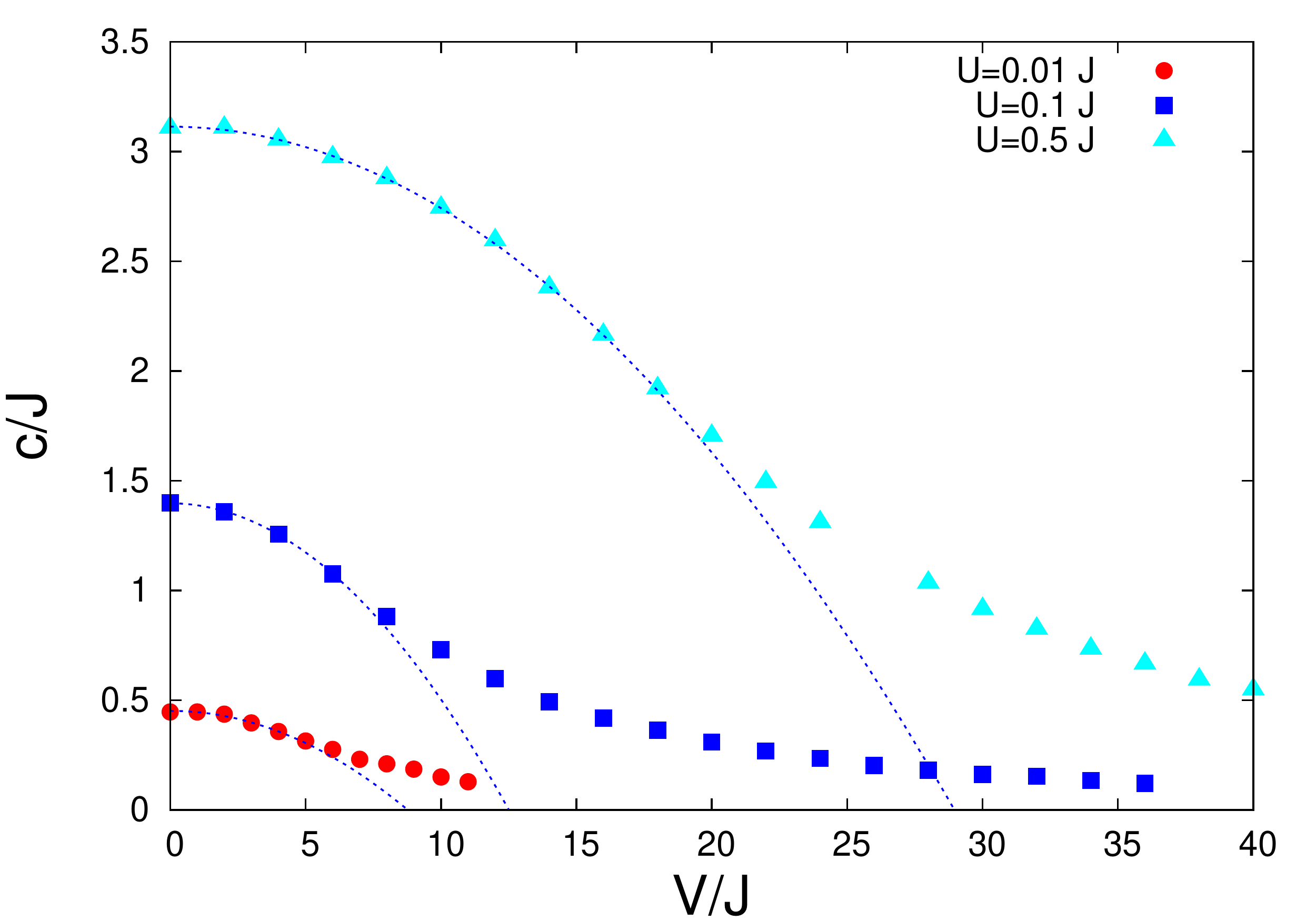}
\caption{Sound velocity as a function of the QP potential strength and
  for various values of the inter-particle repulsion. Other parameters
  as in Fig.~\ref{fig:SqomU0.1}. The dashed curves are parabolic fits
  $c(V) = c(0) - \gamma V^2$ for the lowest $V$ values.}
\label{fig:cBog}
\end{figure}

Even if Bogolyubov theory does not allow us to describe the localized
Bose glass phase for the ground state of the system under
investigation, it still reveals a dramatic evolution in the properties
of the excitations, and a very peculiar nature of the persistent
quasi-condensate phase protected by the interactions. The evolution of
the dynamic structure factor for an increasing strength of the QP
potential is shown in Fig.~\ref{fig:SqomU0.1}. When comparing it to
the non-interacting case of Fig.~\ref{fig:SqomU0}, one clearly
observes substantial analogies. In particular under the effect of the
QP potential, Bogolyubov modes are still organized in sub-bands, which
exhibit sharp dispersion relations in the $(k,\omega)$ plane for
sufficiently weak $V$, while they lose completely their definition in
momentum space when the modes undergo localization for a larger value
of $V$. We observe that the modes at \emph{higher} energy localize at
a \emph{lower} value of $V$, as we will further elaborate upon in the
following. In particular the lowest sub-band, containing the gapless
excitation modes above the ground state, preserves its dispersive
nature for all the values of the QP potential considered, even if the
bandwidth gradually decreases with $V$ - this is exhibited in
Fig.~\ref{fig:SqomU0.1} for $V=16J$, where a low-energy zoom on
$\Sqw(k,\omega)$ is presented for the two strongest values of $V$
shown in Fig.~\ref{fig:SqomU0.1}. In particular the effective
dispersion relation of the lowest band preserves a linear behavior for
$k\to 0$, characteristic of a delocalized sound mode. From the slope
at $k\to 0$, we extract an effective sound velocity $c$, which is
shown in Fig.~\ref{fig:cBog}. We find that $c$ decreases as $c(V)
\approx c(0) - \gamma(U,n)~ V^2$ where $\gamma$ is a constant; this is
consistent with the perturbative results of
Ref.~\cite{Gauletal09,Gauletal11}, showing that the quadratic
dependence on $V$ is a generic property of Bogolyubov modes in the
presence of an external scattering potential.

\begin{figure}[t]
\centering
\includegraphics[width=\textwidth,clip]{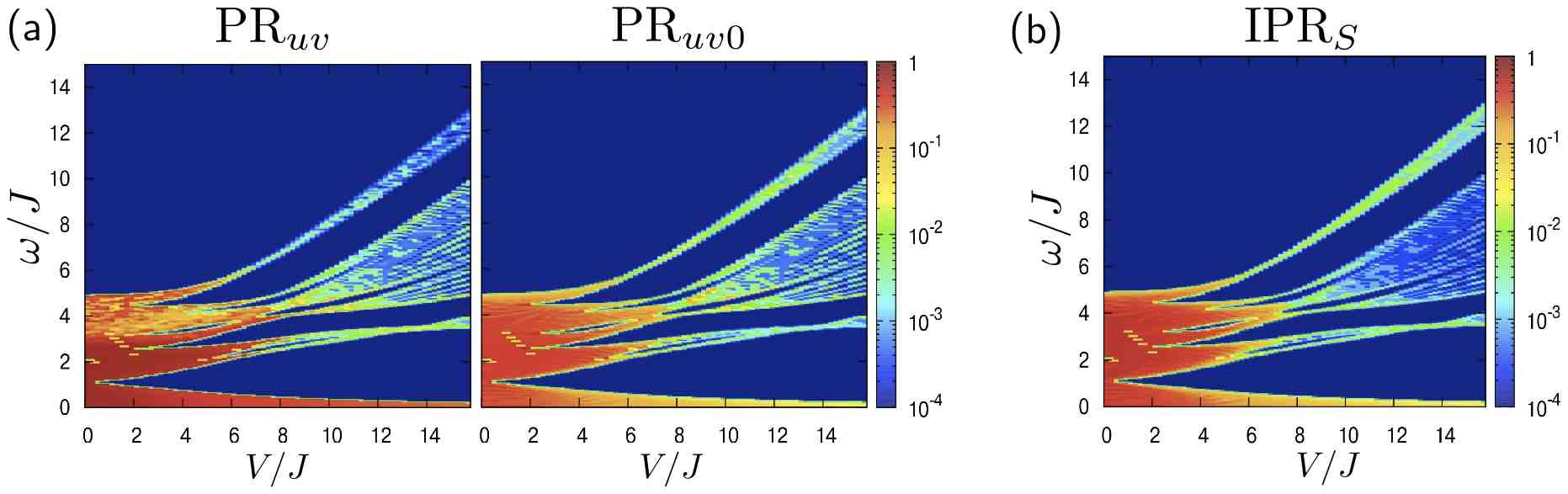}
\caption{(a) Participation ratios PR$_{uv}$ and PR$_{uv0}$ (see text
  for the definition) of variable $V$ and $\omega$. Here $U=0.1$, $r =
  198/256$ and $L=256$; the data are referred to one single
  realization of the phase of the QP potential. The energy axis is
  discretized into intervals of width $\delta\omega = 0.02 J$, and the
  participation ratios (PRs) shown in the figure are an average of the
  values of the PR for the various modes falling within the energy
  interval. The dark-blue regions - associated with a vanishing PR -
  correspond to the gaps between Bogolyubov bands. (b) Inverse
  participation ratio of the dynamic structure factor; same parameters as in the previous panel.}
\label{fig:PR}
\end{figure}

The progressive localization of Bogolyubov modes from the higher to
the lower energies upon increasing the QP potential can be
quantitatively captured by inspecting the effective spatial support of
the $u$ and $v$ lattice functions.
Following the natural definition of norm for the $u$, $v$ functions as
in Eq.~(\ref{e.norm}), and in analogy with the case of normalized
wave functions, one can define a participation ratio (PR) for the $u$,
$v$ functions in the form
\begin{equation}
 {\rm PR}_{uv}(\omega) =  \frac{1}{L} \frac{\left[\sum_j \left( u_{s,j}^2 - v_{s,j}^2 \right)\right]^2}{\sum_j \left( u_{s,j}^2 - v_{s,j}^2 \right)^2}  = 
 \frac{1}{L} \frac{1}{\sum_j \left( u_{s,j}^2 - v_{s,j}^2 \right)^2}~.
\label{e.PRuv}
\end{equation}
This quantity captures the fraction of the system size over which the
$u$, $v$ functions have a non-negligible value. Fig.~\ref{fig:PR}(a)
shows the evolution of PR$_{uv}$ as a function of both QP potential
strength and excitation energy: it is clear that for every finite
value of $V$ the high-energy Bogolyubov modes are more localized than
the low-energy ones, and in particular for $V\sim 6 J$ the highest
band of Bogolyubov modes undergoes localization, and the lower bands
follow in the localization cascade at higher $V$, while the lowest
band remains delocalized over the entire range of $V$ values covered
by the figure.
 
Coming back to the dynamic structure factor, its
expression, Eq.~(\ref{e.SkomBog}), probes the spatial structure of the
overlap function $\sqrt{\rho_j} (u_{s,j} +v_{s,j})$ - giving the local
overlap between the quasi-condensate mode and the excitation mode -
and not simply the spatial structure of the $u$, $v$
functions. Nonetheless, if the condensate mode is delocalized, then
the overlap function has the same localization properties as the $u$,
$v$ functions. This can be directly inspected by plotting the
participation ratio for the overlap function:
\begin{equation}
{\rm PR}_{uv0}(\omega) =  \frac{1}{L} \frac{\left[\sum_j  \sqrt{\rho_j}\left( u_{s,j}+ v_{s,j} \right)\right]^2}{\sum_j  \rho_j \left( u_{s,j}+ v_{s,j} \right)^2}\,.
\end{equation}
As shown in Fig.~\ref{fig:PR}(a), the behavior of this quantity (as a
function of QP potential strength and mode energy) is qualitatively
very similar to the participation ratio of the $u$, $v$ functions of
Eq.~\eqref{e.PRuv}. As a consequence, the dynamic structure factor can
capture the localization properties of the $u$, $v$ functions given
that in its expression, Eq.~\eqref{e.SkomBog}, the Bogolyubov modes
at energy $\epsilon_s$ are weighted by the power spectrum
$|\delta{\tilde\rho}_s(k)|^2$ of the overlap function.  In particular
the power spectrum has opposite localization properties with respect
to the overlap function, namely it is delocalized in $k$ space when
the overlap function is localized and vice versa. Therefore, it appears
natural that the localization properties of the overlap function can
be extracted from the dynamic structure factor by examining its
\emph{inverse} participation ratio (IPR) of in $k$ space
\begin{equation}
{\rm IPR}_S(\omega) = \frac{ \sum_k S^2(k,\omega)}{\Sqw^2(\omega)}~. 
\end{equation} 
In particular one can easily show that $\Sqw(\omega) = L \sum_j
\rho_j \left( u_{s,j}+ v_{s,j} \right)^2$, so that IPR$_S$ and
PR$_{uv0}$ share the same denominator. Fig.~\ref{fig:PR}(b) shows
IPR$_S$ as a function of $V$ and $\omega$; a comparison with the $(V,
\omega)$ dependence of PR$_{uv0}$ shows
striking similarities, demonstrating that the dynamic structure factor
allows to measure directly the localization properties of the
excitation modes in the Bogolyubov regime.
 
\subsection{Results for hardcore bosons}
\label{s.hcbosons}

\subsubsection{Weak QP potential.}

\begin{figure}[t]
\centering
\includegraphics[width=1.05\textwidth,clip]{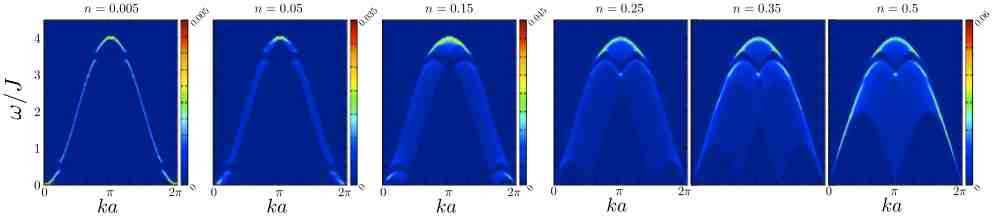}
\caption{Dynamic structure factor for hardcore bosons in a QP
  potential with strength $V=0.25 J$. Here we consider an open chain
  with $L=200$ and variable density, and a single realization of the
  QP potential.}
\label{fig:SkomV0.25}
\end{figure}

For a weak QP potential the (pseudo-)dispersion relation of single
particles is altered as discussed in Section \ref{s.U0}, with the
opening of gaps at $k_{\rm QP}/2$, $\pi\pm k_{\rm QP}/2$, etc... In
Fig.~\ref{fig:SkomV0.25} we show the dynamic structure factor for
hardcore bosons of variable density in a weak QP potential $V=0.25 J$.
In one-dimensional free fermions, some of the dominant features in the
structure factor are related to the transitions between the states at
the bottom of the dispersion relation and states anywhere else in the
energy spectrum - this is due to the singular contribution of the
low-energy states, associated with the van-Hove singularity in their
density of states. As a consequence we observe all single-particle
gaps in the dynamic structure factor as long as Pauli principle allows
the corresponding transitions, namely as long as the Fermi
wave vector, $k_F = \pi n$, is lower than the wave vector of the
arrival state - otherwise the transition is forbidden by Pauli
blocking. Indeed we see that the gap at $k \approx k_{\rm QP}/2$
disappears when $k_F > k_{\rm QP}/2$ (corresponding here to $n \gtrsim
0.23$), and the low-$k$ and low-$\omega$ structure factor is dominated
by the linear mode with dispersion $2J \sin ( k_F ) ~k$.  In what
follows we will focus on this situation, and investigate the case of
(local) filling $n \approx 0.3$.

\subsubsection{Strong QP potential and localization transition.}

\begin{figure}[t]
\centering
\includegraphics[width=1.05\textwidth,clip]{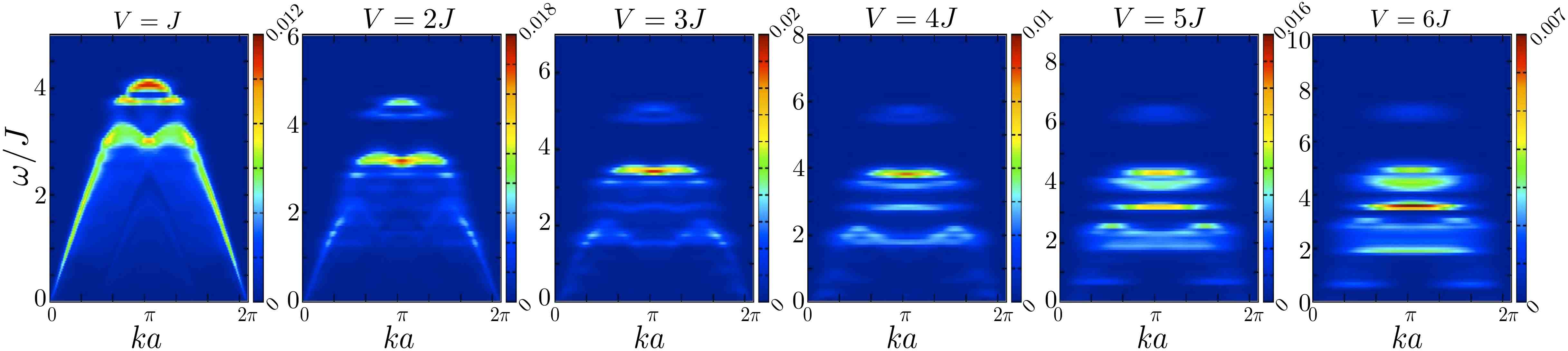}
\caption{$T=0$ dynamic structure factor of 1D hardcore bosons in a
  QP potential with variable strength, and in the presence of a
  confining parabolic potential.}
\label{fig:SqomHCBT0}
\end{figure}

In the presence of a QP potential with strength $V>4 J$ all
single-particle states localize, and therefore we expect a radical
change in the $k$ dependence of the form factors
$\rho_{\alpha\beta}(k)$ in Eq.~\ref{e.SkomHCB}, as the overlap
functions $\psi_{\alpha j}^{*} \psi_{\beta j}$ evolve from extended to
localized. This is indeed observed in Fig.~\ref{fig:SqomHCBT0} where
we consider the evolution of the dynamic structure factor across the
localization transition at $T=0$ for a system of $N=60$ hardcore
bosons in a QP potential, and further confined by a weak harmonic
potential $V_t r_j^2$ (to make contact with an experimentally
realistic situation) with $V_t = 10^{-3} J$. We
observe that the dispersive nature of the excitation modes is quickly
lost as $V$ approaches the critical value, and that the structure
factor fragments into horizontal ridges, namely excitation modes which
are well defined in energy but poorly defined in momentum space. Such
features correspond predominantly to localized particle-hole
excitations, in which the two states $\psi_{\alpha}$ and
$\psi_{\beta}$ connected by the transition are both localized in the
same region of the system, giving a sizable overlap function.

\subsection{Exact diagonalization results: competition between Mott insulator and Bose glass, and comparison between QP and RB potentials}
\label{s.ED}

\begin{figure}[t]
\centering
\includegraphics[width=\textwidth,clip]{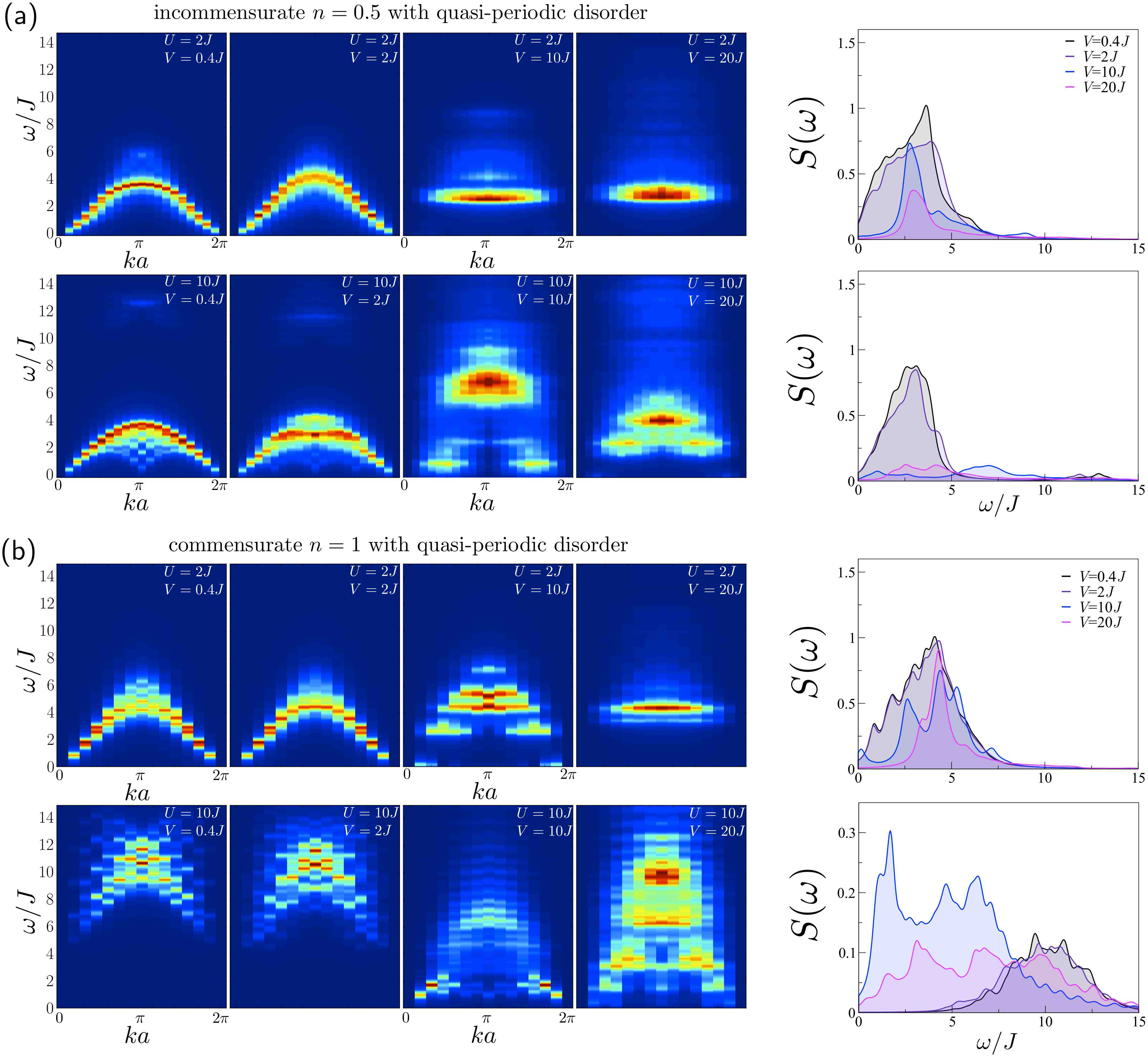}
\caption{Dynamic density structure factor for the bichromatic
  system. \textsf{(a)} for a typical incommensurate density $n=0.5$
  ($L=20$). \textsf{(b)} for the commensurate density $n=1$ ($L=14$)
  where the Mott phase emerges at large $U/J$.}
\label{fig:bichromatic-system}
\end{figure}

\subsubsection{Exact diagonalization results for QP potentials}

We now turn to the results of exact diagonalization in the QP case.
The dynamic structure factor is given in
Fig.~\ref{fig:bichromatic-system} for four typical situations with
increasing QP potential: incommensurate with weak and strong
interactions (Fig.~\ref{fig:bichromatic-system}(a)), and commensurate for the same interaction regimes
(Fig.~\ref{fig:bichromatic-system}(b)).  In the incommensurate case, the effect of the bichromatic
potential is rather weak at $U=2J$ and $V=2J$. The $\omega$ width
increases and the doublon mode is hardly visible but one does not see
the opening of gaps expected from Bogolyubov theory, possibly because
of finite size effects.  Increasing further the potential leads to
localization of the excitations and a spectrum qualitatively similar
to the one predicted using the Bogolyubov approach in
Sec.~\ref{s.Bogo.int}. Notice that although the spectrum apparently
looks gapped due to a large spectral weight for excitations at $\omega
\sim 3J$, it is in fact gapless but with small weights for low-energy
excited states.  The effect of the QP potential is more evident when
starting from $U=10J$ (close to the HCB limit) and increasing
$V$. Subbands in the spectrum do appear (see the panel with $U=10J$,
$V=2J$ in Fig.~\ref{fig:bichromatic-system}) while the doublon mode
loses some of its dispersive features, consistently with the fact that
it can be localized already at weak QP potential strength -- it has a
reduced effective hopping $\sim J^2/U$ -- and its energy is lowered by
the disorder. For a strong QP potential ($V=U$), the spectrum exhibits
many subgaps and it strongly broadens in $k$, in a very similar manner
to what seen in the HCB results of Sec.\ref{s.hcbosons}. The doublon
mode is no longer resolved and the spectrum has large weights over a
broad range of frequencies, while the integrated weight $\Sqw(\omega)$
shows a strong suppression.

Turning to the commensurate case in the superfluid regime ($U=2J$),
here again a sufficiently large bichromatic potential is required to
change the spectrum. Above the localization transition, the spectrum
displays many subgaps typical of the band-folding localization
mechanism (as seen in the panel for $U=2J$, $V=10J$ in
Fig.~\ref{fig:bichromatic-system}(b)) before reaching similar strongly
localized spectrum as in the incommensurate density at very large $V$
(see the $V=20J$ panel). Starting from the MI phase at $U=10 J$, and
introducing a weak QP potential ($V=2J$ in
Fig.~\ref{fig:bichromatic-system}(b)), we observe that the spectral
gap is initially lowered, and mini-gaps appears in the particle-hole
dispersion. For a stronger QP potential ($V=U=10J$ in
Fig.~\ref{fig:bichromatic-system}(b)), the gap closes and the system
enters the strongly-correlated Bose glass phase. The spectrum exhibits
both low-energy excitations with weights around $k=0$ - corresponding
to phonon-like modes of locally superfluid regions - and excitations
at relatively high energies - corresponding to short-wavelength
localized excitations. Increasing further the QP potential to largely
exceed the Mott gap ($V=20 J = 2U$ in
Fig.~\ref{fig:bichromatic-system}(b)), the spectrum appears as
composed of two parts: a low-energy part, associated with regions
exhibiting locally incommensurate densities, and hence a similar
behavior to that of the incommensurately filled lattice; and a
higher-energy part with $\omega \sim U$, associated with localized
particle-hole excitations appearing in regions with local Mott
behavior at commensurate filling.  A similar separation emerges with
the RB distribution as we will see in the following section.  As an intermediate
conclusion, one can keep in mind that the typical signature of the localization due to the QP potential is best
observed when the spectrum possesses subbands and is very broad in
$k$. The spectra for $U=2J$, $V=10J$, \emph{e.g.}, displays nicely
this fingerprint.

\subsubsection{Comparison with random box disorder}

We now compare the effect of a QP potential seen in the previous
section with the effect of true disorder, represented by the RB
distribution. We show results for a RB potential having the exact same
values of the strength $V$ as those discussed for the QP potential.
As already discussed in Ref.~\cite{Roux2008}, the Fourier
  transform of single-particle excitations for a RB potential differs
significantly from those for a QP potential. In particular the gaps
occurring for a QP potential are absent for a RB potential, and
momentum broadening due to localization occurs at an infinitesimal
strength of the RB potential, while a weak QP potential rather leads
to excitations with narrow momentum features, and with a quasi-period
imposed by the pseudo-Brillouin zone of the QP potential. This is due
to the fact that a RB potential scatters Bloch waves at any
wave vector, while a (weak) QP potential primarily affects Bloch waves
with a wave vector $k \approx k_{\rm QP}/2, \pi \pm k_{\rm QP}/2$,
etc. On the contrary, a strong QP potential induces a stronger
  localization than the RB, leading to very broad momentum features.
These differences between the RB case and the QP one will clearly
persist in the Bogolyubov and HCB regimes. In what follows we check
that similar differences are also present for interaction strengths
interpolating between the weakly and strongly interacting regimes.

\begin{figure}[t]
\centering
\includegraphics[width=\textwidth,clip]{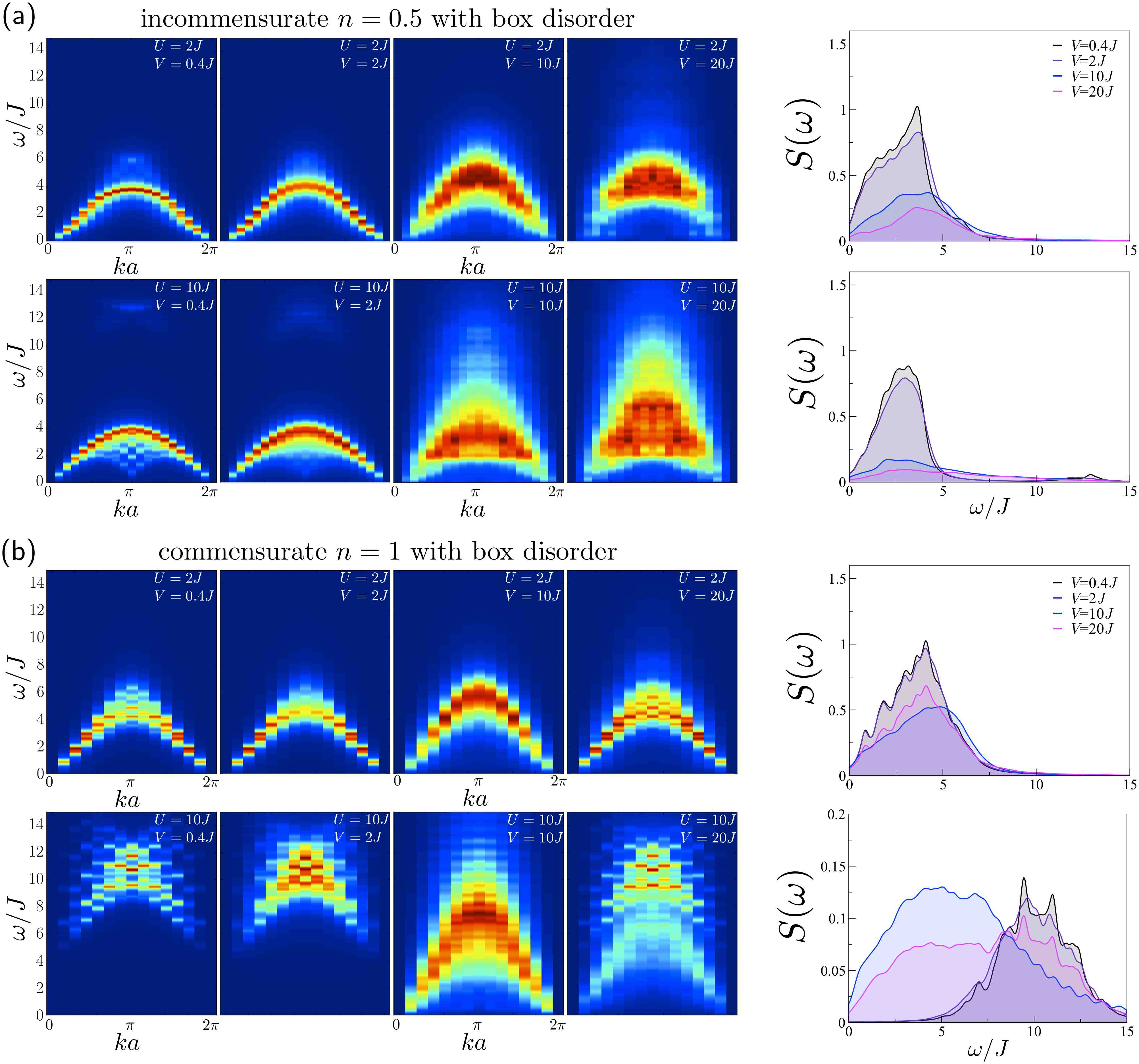}
\caption{Dynamic density structure factor for the RB disordered
  system. \textsf{(a)} for a typical incommensurate density $n=0.5$
  ($L=20$). \textsf{(b)} for the commensurate density $n=1$ ($L=14$).}
\label{fig:disorder-system}
\end{figure}

The results for $S(k,\omega)$ in the presence of a RB potential are
plotted in Fig.~\ref{fig:disorder-system}. The values for the
interaction and disorder strengths are the same as in
Fig~\ref{fig:bichromatic-system}. To avoid repetitions with respect to
the discussion of the QP case, we simply highlight the main analogies
and differences between the RB and QP case. For the incommensurate
density $n=0.5$ and weak interaction $U=2J$, we generally observe that
an arc-like shape of the $S(k,\omega)$ support, typical of the
Bogolyubov mode in the clean case, is preserved in the presence of
disorder, but the energy and momentum structure is strongly broadened,
due to localization of the modes and to the random distribution of
energies of localized excitations induced by disorder.  This is in
sharp contrast with the subband formation seen in the QP potential.
The doublon excitation undergoes a similar fate as for the QP
potential, merging rapidly with the acoustic modes for a large enough
RB potential. For a stronger repulsion $U=10J$ the evolution of
$S(k,\omega)$ with increasing disorder is comparable to the
  quasi-periodic case, certainly due to the fact that finite-size
  effects make the small differences hardly visible. Yet, in the large
  disorder limit, the spectrum is quite different from the QP, with
  the absence of subbands and a weight broadly distributed in
  frequency. In both cases, the spectral weight tends to decrease with
  $V$.

Turning to the commensurate case $n=1$ at weak interaction $U=2J$, we
find that the spectra at weak disorder are comparable to the QP ones,
the disorder leading just to broadening of the dispersion, while the
spectra at strong RB disorder are remarkably different from the QP
case, illustrating the different localization mechanism in the two
cases.  In the case of strong interactions, $U=10 J$, we observe that
the RB potential leads to a closing of the Mott gap, similarly to the
QP potential, but deep in the Bose-glass phase ($V=20 J$),
$S(k,\omega)$ exhibits a rather special structure with two coexisting
features: a low-energy arc-shaped part, quantitatively consistent with
the incommensurate filling case at the same disorder strength (compare
the picture at $U=2J$), and a high-energy part with the same structure
as the particle-hole excitations of the Mott insulator at weak
disorder (compare the case $V=0.4 J$).  As for the QP potential, this
is a clear signature of the strongly-correlated Bose glass regime,
with the coexistence of regions with locally incommensurate filling
and gapless excitations, and regions which preserve a commensurate
filling and a Mott-like behavior -- as also seen (albeit less clearly)
in the case of the QP potential for the same interaction and potential
strength.

An important conclusion of this section is that, the comparison between QP and RB
results at strong disorder shows that Bragg spectroscopy could
certainly probe the very nature of the Bose glass phase and unveil the
localization mechanism at play.

\section{Experimental considerations: finite entropy and $\omega$-scan overlaps}
\label{s.experiments}

\begin{figure}[t]
\centering
\includegraphics[width=1.05\textwidth,clip]{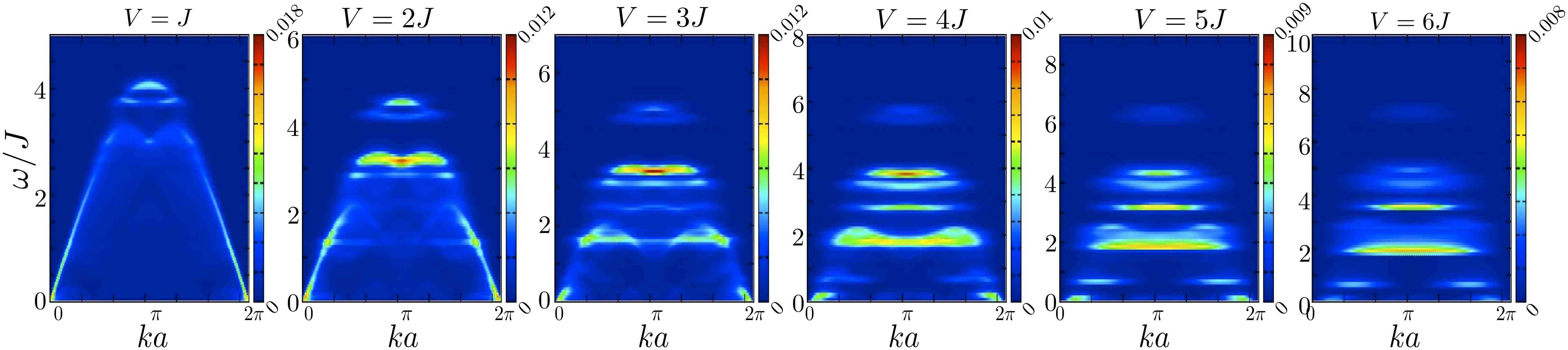}
\caption{Dynamic structure factor of 1D
  bosons in a QP potential with variable strength and in a confining
  parabolic potential with strength $V_t = 10^{-3} J$. Here the QP potential is increased
  adiabatically at fixed entropy per particle ${\cal S}/N =$ 1 k$_B$.}
\label{fig:SqomHCBS1}
\end{figure}

In the previous section we have seen that low-energy features of the
spectral function at large $U$ reproduce the HCB behavior. Therefore,
one can exploit the exact solution available for the HCB case to make
further contact with an experimentally realistic situation. Given that
in experiments the loading of the optical lattices occurs in a
(quasi)-adiabatic way, we study the evolution of the dynamic structure
factor with an increasing height of the secondary lattice at fixed
entropy per particle, taken to be ${\cal S}/N = 1\, \text{k}_B$. The
corresponding temperature which enters in the calculation of the
dynamic structure factor in Eq.~(\ref{e.SkomHCB}) is obtained by
numerical inversion of the equation which links temperature and
entropy for free fermions
\begin{equation}
{\cal S}(T) = -{\rm k}_B \sum_\alpha \left[ f_{\alpha} \log f_{\alpha} +  (1- f_{\alpha}) \log (1-f_{\alpha}) \right]
\end{equation}
where $f_{\alpha} = f(e_{\alpha},T)$.  In the present case, since the
QP potential reduces the density of states at low energy and fragments
the energy spectrum into increasingly spaced minibands, entropy
conservation implies adiabatic heating of the system as $V$ increases.
Yet, the comparison between Figs.~\ref{fig:SqomHCBT0}
  and~\ref{fig:SqomHCBS1} shows that the main features of the
localization transition remain intact, indicating that they are
accessible to current experiments.

As a second aspect with direct experimental relevance, we propose an
effective, global way to capture the dispersive or non-dispersive
(namely $k$-dependent or $k$-independent) nature of the dynamic
structure factor. This amounts  to considering the
\emph{overlap} function of two $\omega$-scans in $\Sqw(k,\omega)$ at
wave vectors $k$ and $k+\Delta k$:
\begin{equation}
 O(k, k+\Delta k) = \frac{\displaystyle \int d\omega ~\Sqw(k,\omega)  \Sqw(k+\Delta k,\omega)}{\displaystyle \left[ \int d\omega  ~S^2(k,\omega) ~\int d\omega  ~S^2(k+\Delta k,\omega) \right]^{1/2}} ~.
\end{equation}
This expression is normalized so that $O(k,k) = 1$, and in general
$O(k,k+\Delta k)$ is close to 1 if the features in $\Sqw(k,\omega)$ at
$k$ and $k+\Delta k$ have a large overlap; if on the other hand
$\Sqw(k,\omega)$ has strongly dispersive features, $O$ is in general
$\ll 1$ - it exactly vanishes in the extreme limit of a
$\Sqw(k,\omega) \sim \delta(\omega-\epsilon_k)$, displaying a
$\delta$-ridge associated with a dispersion relation $\epsilon_k$
having a finite group velocity around $k$, $d\epsilon_k/dk \neq 0$.

\begin{figure}[t]
\centering
\includegraphics[width=\textwidth,clip]{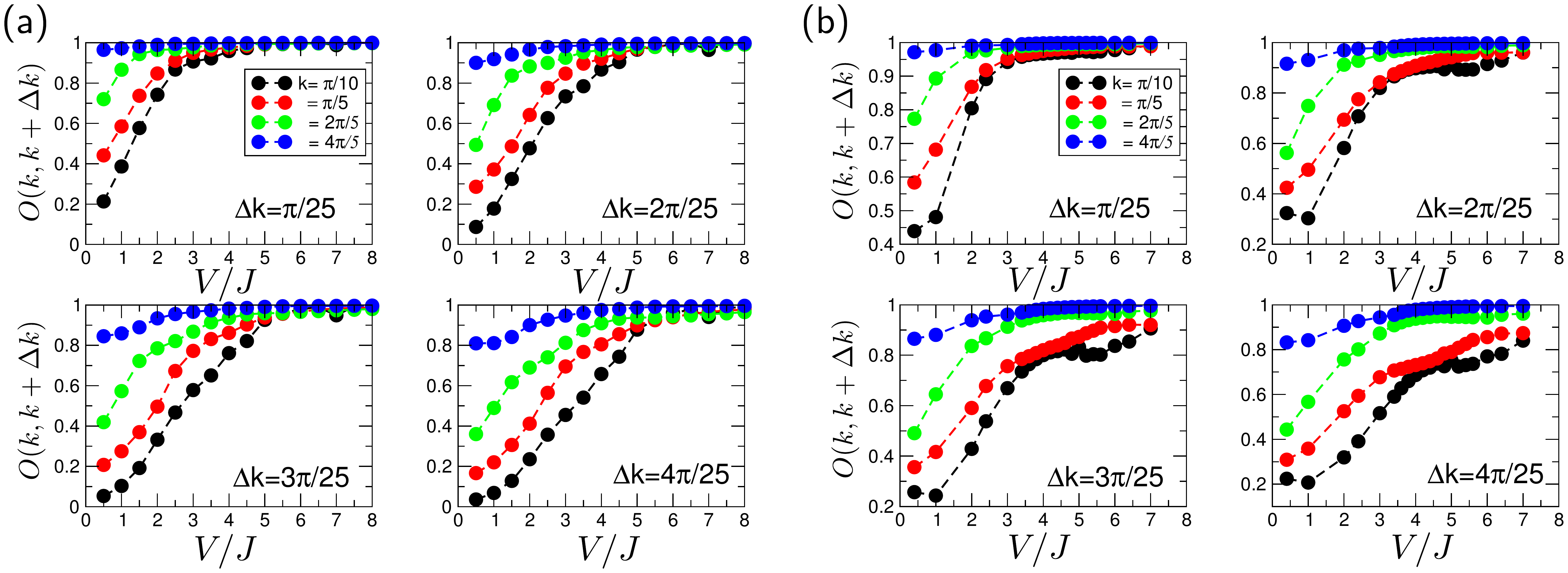}
\caption{(a) Overlap function of $\omega$-scans in the dynamic
  structure factor for $N=60$ hardcore bosons at $T=0$ and in a QP
  potential plus a confining parabolic potential. Each panel shows the
  overlap at four different reference wave vectors $k$, with an
  increasing separation $\Delta k$ when going from the
  upper-left to the lower-right panel. (b) Same but for finite
  entropy, ${\cal S}/N =$ 1 k$_B$.}
\label{fig:overlapT0}
\end{figure}

In Figs.~\ref{fig:overlapT0}(a-b) we show the overlap function at zero
and finite entropy (1 k$_B$ per particle) respectively, for four
different values of $k$ (going from the Brillouin zone center towards
the edge) and for four values of the separation $\Delta k$ between
$\omega$-scans, as a function of the QP potential strength $V$. For
all scans we observe that the overlap increases for $V/J\to 4^{-}$,
and it displays plateaus in the localized phase; this feature is not
only robust to the presence of a confining potential, but also to the
presence of a finite entropy. In particular we observe that the
overlap is most sensitive to the localization transition when $k$ is
close to the zone center, because this is the wave-vector region in
which the features in $\Sqw(k,\omega)$ display the strongest $k$
dependence in the delocalized phase. Unsurprisingly, the sharpest
features in $O$ at the localization transition are displayed for the
largest $\Delta k$ separation we investigated.  Therefore the overlap
provides a quantitative estimate of the localization transition based
on the dynamic structure factor. Its effectiveness relies upon the
fact that, in the problem at hand, \emph{all} excitation modes
localize and lose their dispersive nature at the ground-state
transition point $V=4J$. Yet it can be extended to situations (as the
one discussed in Section \ref{s.Bogo}) in which only a part of the
excited states in the excitation spectrum localizes. Most importantly,
the overlap is a convenient experimental observable which allows to
detect the transition even when one does not have access experimentally to
the whole $(k, \omega)$ plane. Indeed several recent experiments
\cite{Veeravalli2008, Clement2009, Fabbri2011, Fabbri2012} have
measured $\omega$-scans for a fixed $k$ or a small range of $k$ values
(limited by the optical access to the sample). We observe that the
overlap is well adapted to such a situation, providing a direct
inspection into the localization transition by using only two
tomographic scans of the dynamic structure factor at near wave-vectors.

\section{Summary and conclusions}
\label{s.conclusions}

In conclusion, we have shown that the study of the dynamic structure
factor provides important information on the interplay of disorder and
interactions in one dimensional Bose fluids. Our analysis relies upon
exact diagonalization results at arbitrary interactions, Bogolyubov
theory for the weakly interacting case, and the exact solution for the
hardcore case. Already in the clean case, the dynamic structure factor
displays different features in the various interaction regimes, from
sharp dispersive features in the weakly interacting superfluid phase,
to a broad particle-hole continuum in the strongly-interacting (but
still superfluid) phase, coexisting with a doublon mode. Once the
disorder is turned on, we have shown that the dynamic structure
factor allows to explore the spatial support of the excitations,
yielding information on their localization properties. We have also
investigated the features of the dynamic structure factor in the
presence of a random-box disorder, from weak to strong interactions,
showing that the dynamic structure factor captures the differences in
the spectral features of the excitations with respect to the
quasi-periodic potential. Thereby, it can probe in a direct way the
localization mechanism. Finally, by exploiting the exact solution
available for the hardcore-boson limit, we have shown that the main
features of localization exhibited by the dynamic structure factor at
zero temperature remain visible at a realistic finite entropy; and we
have suggested an experimentally viable method to extract information
on the Bose-glass transition at strong interactions, by analyzing
frequency scans in the dynamic structure factor at two fixed and near
wave vectors. While our paper exclusively focuses on the
one-dimensional case, we argue that the insight provided by the
dynamic structure factor into the physics of disordered bosons will be
extremely useful also in higher-dimensional cases. In particular, the
sensitivity of the dynamic structure factor to the extended or
localized nature of the excitation modes, in a given frequency range,
makes it a most viable probe of the presence of a mobility edge in the
spectrum, which is a characteristic feature of higher-dimensional
systems~\cite{DeMarco, Jendrzejewski2012}.

\subsection{Acknowledgments}

Useful discussions with N. Fabbri, L. Fallani, C. Fort, and E. Orignac
are gratefully acknowledged.  GR acknowledges support from the Agence
Nationale de la Recherche under grant ANR-2011-BS040-012-01 QuDec and
AM from the ERC grant "Handy-Q" No. 258608. Numerical calculations were partially performed on the computer cluster of the PSMN (ENS Lyon). 

\section*{References}

\bibliographystyle{prsty}
\bibliography{modulation}

\begin{thebibliography}{10}

\bibitem{Sanchez-Palencia2010}
L. Sanchez-Palencia and M. Lewenstein, Nature Physics {\bf 6},  87  (2010).

\bibitem{Aspect}
J. Billy {\it et~al.}, Nature {\bf 453},  891  (2008).

\bibitem{Pasienskieta10}
M. Pasienski {\it et~al.}, Nat. Phys. {\bf 6},  677  (2010).

\bibitem{Robert-de-Saint-Vincent2010}
M. Robert-de Saint-Vincent {\it et~al.}, Phys. Rev. Lett. {\bf 104},  220602
  (2010).

\bibitem{DeMarco}
S.~S. Kondov, W.~R. McGehee, J. Zirbel, and B. DeMarco, Science {\bf 334},  66
  (2011).

\bibitem{Jendrzejewski2012}
F. {Jendrzejewski} {\it et~al.}, Nature Physics {\bf 8},  398  (2012).

\bibitem{Jendrzejewski2012a}
F. Jendrzejewski {\it et~al.}, Phys. Rev. Lett. {\bf 109},  195302  (2012).

\bibitem{Fallani}
L. Fallani {\it et~al.}, Phys. Rev. Lett. {\bf 98},  130404  (2007).

\bibitem{Fallani2006}
J.~E. Lye {\it et~al.}, Phys. Rev. A {\bf 75},  061603  (2007).

\bibitem{Guarrera2007}
V. Guarrera {\it et~al.}, New J. Phys. {\bf 9},  107  (2007).

\bibitem{Guarreraetal08}
V. Guarrera {\it et~al.}, Phys. Rev. Lett. {\bf 100},  250403  (2008).

\bibitem{LENS}
G. Roati {\it et~al.}, Nature {\bf 453},  895  (2008).

\bibitem{Deissler2009}
B. Deissler {\it et~al.}, Nat. Phys. {\bf 6},  354  (2010).

\bibitem{Deissleretal11}
B. Deissler {\it et~al.}, New Journal of Physics {\bf 13},  023020  (2011).

\bibitem{Lucionietal11}
E. Lucioni {\it et~al.}, Phys. Rev. Lett. {\bf 106},  230403  (2011).

\bibitem{AubryA79}
S. Aubry and G. Andr{\'e}, Ann. Israel. Phys. Soc. {\bf 3},  133  (1980).

\bibitem{Giamarchi1987}
T. Giamarchi and H.~J. Schulz, Europhys. Lett. {\bf 3},  1287  (1987).

\bibitem{Giamarchi1988}
T. Giamarchi and H.~J. Schulz, Phys. Rev. B {\bf 37},  325  (1988).

\bibitem{Fisher1989}
M.~P.~A. Fisher, P.~B. Weichman, G. Grinstein, and D.~S. Fisher, Phys. Rev. B
  {\bf 40},  546  (1989).

\bibitem{OzeriRMP}
R. Ozeri, N. Katz, J. Steinhauer, and N. Davidson, Rev. Mod. Phys. {\bf 77},
  187  (2005).

\bibitem{Menotti2003}
C. Menotti, M. Kr\"amer, L. Pitaevskii, and S. Stringari, Phys. Rev. A {\bf
  67},  053609  (2003).

\bibitem{CauxCalabrese}
J.-S. Caux and P. Calabrese, Phys. Rev. A {\bf 74},  031605  (2006).

\bibitem{Caux2007}
J.-S. Caux, P. Calabrese, and N.~A. Slavnov, J. Stat. Mech. {\bf 2007},  P01008
   (2007).

\bibitem{Calabrese2007}
P. Calabrese and J.-S. Caux, Phys. Rev. Lett. {\bf 98},  150403  (2007).

\bibitem{Orignac2012}
E. Orignac, R. Citro, S. De~Palo, and M.-L. Chiofalo, Phys. Rev. A {\bf 85},
  013634  (2012).

\bibitem{Huber2007}
S.~D. Huber, E. Altman, H.~P. B\"{u}chler, and G. Blatter, Phys. Rev. B {\bf
  75},  085106  (2007).

\bibitem{Golovach2009}
V.~N. Golovach, A. Minguzzi, and L.~I. Glazman, Phys. Rev. A {\bf 80},  043611
  (2009).

\bibitem{Grass2011}
T.~D. Gra\ss{}, F.~E.~A. dos Santos, and A. Pelster, Phys. Rev. A {\bf 84},
  013613  (2011).

\bibitem{Roth2004}
R. Roth and K. Burnett, J. Phys. B: At. Mol. Opt. Phys. {\bf 37},  3893
  (2004).

\bibitem{Batrouni2005}
G.~G. Batrouni, F.~F. Assaad, R.~T. Scalettar, and P.~J.~H. Denteneer, Phys.
  Rev. A {\bf 72},  031601  (2005).

\bibitem{Rey2005}
A.~M. Rey {\it et~al.}, Phys. Rev. A {\bf 72},  023407  (2005).

\bibitem{Pupillo2006}
G. Pupillo, A.~M. Rey, and G.~G. Batrouni, Phys. Rev. A {\bf 74},  013601
  (2006).

\bibitem{Pippan2009}
P. Pippan, H.~G. Evertz, and M. Hohenadler, Phys. Rev. A {\bf 80},  033612
  (2009).

\bibitem{Ejima2011}
S. Ejima, H. Fehske, and F. Gebhard, Europhys. Lett. {\bf 93},  30002  (2011).

\bibitem{Ejima2012}
S. Ejima {\it et~al.}, Phys. Rev. A {\bf 85},  053644  (2012).

\bibitem{Clement2009}
D. Cl\'{e}ment {\it et~al.}, Phys. Rev. Lett. {\bf 102},  155301  (2009).

\bibitem{Du2010}
X. Du {\it et~al.}, New Journal of Physics {\bf 12},  083025  (2010).

\bibitem{Ernst2010}
P. Ernst {\it et~al.}, Nature Physics {\bf 6},  56  (2010).

\bibitem{Fabbri2011}
N. Fabbri {\it et~al.}, Phys. Rev. A {\bf 83},  031604  (2011).

\bibitem{Fabbri2012}
N. Fabbri {\it et~al.}, Phys. Rev. Lett. {\bf 109},  055301  (2012).

\bibitem{Orso2009}
G. Orso, A. Iucci, M.~A. Cazalilla, and T. Giamarchi, Phys. Rev. A {\bf 80},
  033625  (2009).

\bibitem{Knap2010}
M. Knap, E. Arrigoni, and W. von~der Linden, Phys. Rev. A {\bf 82},  053628
  (2010).

\bibitem{Roux2008}
G. Roux {\it et~al.}, Phys. Rev. A {\bf 78},  023628  (2008).

\bibitem{Batrouni1990}
G.~G. Batrouni, R.~T. Scalettar, and G.~T. Zimanyi, Phys. Rev. Lett. {\bf 65},
  1765  (1990).

\bibitem{Scalettar1991}
R.~T. Scalettar, G.~G. Batrouni, and G.~T. Zimanyi, Phys. Rev. Lett. {\bf 66},
  3144  (1991).

\bibitem{Freericks1996}
J.~K. Freericks and H. Monien, Phys. Rev. B {\bf 53},  2691  (1996).

\bibitem{Prokofev1998}
N.~V. Prokof'ev and B.~V. Svistunov, Phys. Rev. Lett. {\bf 80},  4355  (1998).

\bibitem{Rapsch1999}
S. Rapsch, U. Schollw{\"o}ck, and W. Zwerger, Europhys. Lett. {\bf 46},  559
  (1999).

\bibitem{Simon1982}
B. Simon, Advances in Applied Mathematics {\bf 3},  463   (1982).

\bibitem{Sokoloff85}
J. Sokoloff, Phys. Rep. {\bf 126},  189  (1985).

\bibitem{Thouless1983}
D.~J. Thouless, Phys. Rev. B {\bf 28},  4272  (1983).

\bibitem{Kohmoto1983}
M. Kohmoto, L.~P. Kadanoff, and C. Tang, Phys. Rev. Lett. {\bf 50},  1870
  (1983).

\bibitem{Tang1986}
C. Tang and M. Kohmoto, Phys. Rev. B {\bf 34},  2041  (1986).

\bibitem{Hiramoto1992}
H. Hiramoto and M. Kohmoto, Int. J. Mod. Phys. B {\bf 06},  281  (1992).

\bibitem{Barache1994}
D. Barache and J.~M. Luck, Phys. Rev. B {\bf 49},  15004  (1994).

\bibitem{Albert2010}
M. Albert and P. Leboeuf, Phys. Rev. A {\bf 81},  013614  (2010).

\bibitem{Vidal1999}
J. Vidal, D. Mouhanna, and T. Giamarchi, Phys. Rev. Lett. {\bf 83},  3908
  (1999).

\bibitem{Vidal2001}
J. Vidal, D. Mouhanna, and T. Giamarchi, Phys. Rev. B {\bf 65},  014201
  (2001).

\bibitem{Roth2003}
R. Roth and K. Burnett, Phys. Rev. A {\bf 68},  023604  (2003).

\bibitem{Roscilde2008}
T. Roscilde, Phys. Rev. A {\bf 77},  063605  (2008).

\bibitem{Deng2008}
X. Deng, R. Citro, A. Minguzzi, and E. Orignac, Phys. Rev. A {\bf 78},  013625
  (2008).

\bibitem{PinesNozieres}
D. Pines and P. Nozieres, {\em The theory of quantum liquids} (W.A. Benjamin,
  New York, 1966).

\bibitem{Brunelloetal2001}
A. Brunello {\it et~al.}, Phys. Rev. A {\bf 64},  063614  (2001).

\bibitem{Veeravalli2008}
G. Veeravalli, E. Kuhnle, P. Dyke, and C.~J. Vale, Phys. Rev. Lett. {\bf 101},
  250403  (2008).

\bibitem{Lu2011}
B. Lu {\it et~al.}, Phys. Rev. A {\bf 83},  051608  (2011).

\bibitem{MoraC03}
C. Mora and Y. Castin, Phys. Rev. A {\bf 67},  053615  (2003).

\bibitem{BlaizotR84}
J.-P. Blaizot and G. Ripka, {\em Quantum Theory of Finite Systems} (MIT Press,
  Cambridge, MA (USA), 1986).

\bibitem{Fontanesietal09}
L. Fontanesi, M. Wouters, and V. Savona, Phys. Rev. Lett. {\bf 103},  030403
  (2009).

\bibitem{Fontanesietal10}
L. Fontanesi, M. Wouters, and V. Savona, Phys. Rev. A {\bf 81},  053603
  (2010).

\bibitem{LuganSP11}
P. Lugan and L. Sanchez-Palencia, Phys. Rev. A {\bf 84},  013612  (2011).

\bibitem{GaulM11}
C. Gaul and C.~A. M\"uller, Phys. Rev. A {\bf 83},  063629  (2011).

\bibitem{FontanesiPhD}
L. Fontanesi, {\em PhD Thesis, EPFL} (unpublished, Lausanne, 2011).

\bibitem{WuGriffin97}
W.-C. Wu and A. Griffin, Phys. Rev. A {\bf 54},  4204  (1996).

\bibitem{Zambellietal00}
F. Zambelli, L. Pitaevskii, D.~M. Stamper-Kurn, and S. Stringari, Phys. Rev. A
  {\bf 61},  063608  (2000).

\bibitem{Castroetal06}
E.~V. Castro, N.~M.~R. Peres, K.~S.~D. Beach, and A.~W. Sandvik, Phys. Rev. B
  {\bf 73},  054422  (2006).

\bibitem{LiebSM61}
E. Lieb, T. Schultz, and D. Mattis, Ann. Phys. {\bf 16},  407  (1961).

\bibitem{Winkler2006}
K. {Winkler} {\it et~al.}, Nature {\bf 441},  853  (2006).

\bibitem{Imambekov2012}
A. Imambekov, T. Schmidt, and L. Glazman, Rev. Mod. Phys. {\bf 84},  1253
  (2012).

\bibitem{Cloizeaux1962}
J. des Cloizeaux and J.~J. Pearson, Phys. Rev. {\bf 128},  2131  (1962).

\bibitem{CastinM03}
C. Mora and Y. Castin, Phys. Rev. A {\bf 67},  053615  (2003).

\bibitem{CetoliL10}
A. Cetoli and E. Lundh, Phys. Rev. A {\bf 81},  063635  (2010).

\bibitem{Gauletal09}
C. Gaul, N. Renner, and C.~A. M\"uller, Phys. Rev. A {\bf 80},  053620  (2009).

\bibitem{Gauletal11}
C. Gaul and C.~A. M\"uller, Phys. Rev. A {\bf 83},  063629  (2011).

\end{thebibliography}

\end{document}